\documentclass[12pt,a4paper]{article}
\usepackage{epsfig}
\newcommand{\be}{\begin{equation}}
\newcommand{\ee}{\end{equation}}
\newcommand{\ba}{\begin{eqnarray}}
\newcommand{\ea}{\end{eqnarray}}
\newcommand{\baa}{\begin{eqnarray*}}
\newcommand{\btab}{\begin{tabular}}
\newcommand{\etab}{\end{tabular}}
\newcommand{\eaa}{\end{eqnarray*}}

\newcommand{\bit}{\begin{itemize}}
\newcommand{\eit}{\end{itemize}}

\def\UV{\hbox{\tiny UV}}

\def\Im{\,{\rm Im}\, }

\newcommand\re[1]{(\ref{#1})}

\newcommand\lr[1]{{\left({#1}\right)}}

\catcode`\@=11 
\def \lsim {\mathrel{\mathpalette\@versim<}}
\def \gsim {\mathrel{\mathpalette\@versim>}}
\def\gappeq{\mathrel{\rlap {\raise.5ex\hbox{$>$}}{\lower.5ex\hbox{$\sim$}}}}
\def\lappeq{\mathrel{\rlap{\raise.5ex\hbox{$<$}}{\lower.5ex\hbox{$\sim$}}}}
\def\@versim#1#2{\vcenter{\offinterlineskip
\ialign{$\m@th#1\hfil##\hfil$\crcr#2\crcr\sim\crcr } }}
\catcode`\@=12 
\newcommand \vev [1] {\langle{#1}\rangle}

\newcommand{\mycomm}[1]{\hfill\break $\phantom{a}$\kern-3.5em{\tt===$>$ \bf
#1}\hfill\break}
\newcommand{\mycommA}[1]{\hfill\break $\phantom{a}$\kern-3.5em{\tt***$>$ \bf
#1}\hfill\break}
\renewcommand{\thefootnote}{\fnsymbol{footnote}}
\newcommand{\ysl}{\mbox{$y$\hspace{-0.5em}\raisebox{0.1ex}{$/$}}}

\def \e {\mbox{e}}
\def \CO {{\cal O}}

\def \as {\relax\ifmmode\alpha_s\else{$\alpha_s${ }}\fi}

\usepackage{epsfig}
\title{essai}
\author{}
\date{\today}

\begin{document}

\begin{titlepage}
\begin{flushright}
{\small CERN-TH/2002-060}\\
{\small LPT-Orsay-02-023}\\
{\small SHEP-02/02}\\
{\small March, 2002}

\end{flushright}
\vspace{.13in}

\begin{center}
{\Large {\bf Power corrections in deep inelastic \\
structure functions at large Bjorken~$x$}}\footnote{Research
supported in part by the EC
program ``Training and Mobility of Researchers'', Network
``QCD and Particle Structure'', contract ERBFMRXCT980194.}

\vspace{.4in}

{\bf E.~Gardi} $^{(1)}$, {\bf G.P.~Korchemsky} $^{(2)}$, {\bf D.A.~Ross} $^{(3)}$
and {\bf S.~Tafat} $^{(2)}$

\vspace{0.25in}

$^{(1)}$ TH Division, CERN, CH-1211 Geneva 23, Switzerland\\

$^{(2)}$ Laboratoire de Physique Th\'eorique\footnote{CNRS UMR 8627},
Universit\'e de Paris XI, 91405 Orsay Cedex, France\\

$^{(3)}$ Department of Physics and Astronomy, University of Southampton,
Southampton SO17 1BJ, U.K.

\vspace{.4in}

\end{center}
\noindent  {\bf Abstract:}
We study power corrections to the spin-averaged structure functions~$F_2$
and~$F_L$ in the semi-inclusive region. The dramatic breakdown of the operator
product expansion at large Bjorken~$x$ due to the formation of a narrow jet with invariant mass~$W$ ($W^2=Q^2(1-x)/x$) calls for an alternative approach.
Our main conjecture is that the dominant contribution at each twist is the one which
mixes under renormalization with the leading twist (ultraviolet dominance).
At twist two the dominant perturbative corrections at large $x$  due to soft and collinear radiation can be factorized into a jet function. From the ultraviolet dominance conjecture
it follows that the twist two parton distribution as well as the jet function actually factorize
to any order in the twist expansion. This factorization suggests that non-perturbative corrections~$\sim 1/W^{2n}$ exponentiate together with the leading twist.
We verify explicitly, at the level of a single dressed gluon, the cancellation
between ambiguities due to infrared renormalons in the twist-two coefficient
functions and those due to the ultraviolet divergence of twist-four matrix
elements. Independently of the renormalon analysis, we show that the dominant
contribution to twist four at large~$x$ is associated with a twist-two like
configuration: the final states are identical to those of twist two, whereas the initial states differ just by additional partons carrying small momentum fractions.
This picture is consistent with the ultraviolet dominance conjecture.
\vspace{.25in}
\end{titlepage}

\def\thefootnote{\arabic{footnote}}
\setcounter{footnote} 0

\section{Introduction}

The phenomenological study of the structure functions of deep inelastic
scattering (DIS) is well developed~\cite{MRST} and its role in future colliders
can hardly be overestimated~\cite{Catani}. One of the gaps in the
state-of-the-art QCD description of the structure functions is the region of
large Bjorken $x$, the so-called semi-inclusive
region~\cite{MRST,Olness}. The two limitations are insufficient data
and lack of pure theoretical understanding of this region. The QCD analysis shows
large perturbative and non-perturbative corrections which make the parton model
inapplicable at large $x$.

In general, the Operator Product Expansion
(OPE)~\cite{Jaffe:zw} provides a well-established framework
for analyzing the structure functions in QCD. It allows one to expand the moments
of the structure functions $F_a$, where $a=L,\,2$, in inverse powers of the hard
scale $1/Q^2$,
\be
\int_0^1 dx\,x^{N-2}\,F_a(x,Q^2) =
\sum_{i=q,g}\,\widetilde{C}_{2,a}^{i}(N,\alpha_s)\, \langle O_i^{(2)}(N)\rangle
+ \frac{\sum_{j}\,\widetilde{C}_{4,a}^{j}(N,\alpha_s)\, \langle O_j^{(4)}(N)\rangle}{Q^2}\,+\,\ldots
\label{moments-def}
\ee
expressing each term in the expansion as a product of  the matrix
elements of certain local composite twist$-n$ operators between hadronic states,
$ \langle O_j^{(n)}(N)\rangle$, and the corresponding coefficient functions
$\widetilde{C}_{n,a}^{j}(N,\alpha_s)$. The latter can be systematically calculated in perturbative
QCD, while the former bring a set of universal non-perturbative parameters, corresponding
to generalized moments of some non-perturbative distributions.

The number of the matrix elements entering the expression for the $N$-th moment
of the structure function \re{moments-def} depends on the twist. The leading,
twist-two flavour non-singlet, contribution to~\re{moments-def} depends, for
arbitrary $N$, on just one non-perturbative parameter, the $N$-th moment of quark
distribution within the hadron. The flavour singlet contribution depends on the
$N$-th moment of both the quark and the gluon distributions. In any case,
twist-two depends strictly on {\em single particle distributions}. The asymptotic
behaviour of the structure function at large $x$ corresponds, in moment space, to
large $N\sim1/(1-x)$. An additional simplification occurs in this limit at twist
two: the contribution of gluons to the structure functions becomes suppressed as
compared with that of the quarks \cite{DDKS}--\cite{Zijlstra:1992qd}.

Going over to higher twists, one encounters a completely different situation: the
number of matrix elements entering the twist-$n$ $(n\ge 4)$ contribution depends
on the moment $N$ and it grows rapidly with $N$. The reason is that the
high-twist contribution depends on novel {\em multi-particle distributions},
which describe the correlations between partons in the nucleon and which do not
have a simple partonic interpretation~\cite{JS}--\cite{Balitsky:1989bk}.
For different moments $N$, the high-twist matrix
elements measure specific projections of the multi-parton distributions. One
consequence is that the $Q^2$ dependence of the high-twist contribution to the
structure functions is not described by the DGLAP evolution equations.

Examining the large-$x$ limit one finds that the twist expansion breaks down. The
physical reason for this is that for large-$x$ kinematics the final state in the
current fragmentation region is bound to be a narrow jet with a square invariant
mass $W^2=Q^2(1-x)/x$ such that $W^2\ll Q^2$. In such a situation,  the
long-distance effects of the jet formation dominate, in spite of the momentum
transfer $Q^2$ being large: the twist expansion goes effectively in inverse
powers of the smallest scale,~$1/W^2$. As indicated by recent phenomenological
analysis~\cite{Simula}--\cite{LEKN}, the problem becomes acute when $W^2
\,\lsim {\rm few\,\, GeV}^2$.

The breakdown of the OPE at large $x$ is three-fold: the perturbative expansion breaks down, the power expansion breaks down, and the number
of non-perturbative parameters increases.
Perturbative corrections to the twist-two coefficient functions $\widetilde{C}_{2,a}^{i}(N,\alpha_s)$
is enhanced at large $N$ by Sudakov logarithms
\be
\widetilde{C}_{2,a}^{i}(N,\alpha_s)\sim \sum_{m\ge 0} \alpha_s^m\lr{a_m\ln^{2m}N+ b_m
\ln^{2m-1}N+ \cdots}\,.
\label{Sudak}
\ee
These large corrections originate from an incomplete cancellation between the
contribution of real and virtual soft gluons at large $x$. To obtain a reliable
perturbative description of the coefficient functions, they need to be resummed
to all orders in~$\alpha_s$. Thanks to the factorization property of soft
radiation such resummation can be performed~\cite{Sterman}--\cite{KM} and it takes the form of exponentiation.
One expects to find similar Sudakov corrections in
the coefficient functions of higher twist, $\widetilde{C}_{n,a}^{i}(N,\alpha_s)$ with $n > 2$.
Since higher-twist corrections scale as powers of $1/W^2\sim N/Q^2$ rather than
as $1/Q^2$, terms of arbitrarily high twist in~\re{moments-def} become relevant
at large $N$.
Consequently, the number of matrix elements involved increases sharply.
This means that, without additional insight into the properties of the high
twists, the OPE parameterization of non-perturbative corrections to the structure
functions in the semi-inclusive region becomes impractical and their scale
dependence is obscure.

In the absence of a practical OPE-based parametrization of power corrections to
structure functions, various models were developed~\cite{Beneke}--\cite{Guo} and
applied~\cite{Simula}--\cite{LEKN}. In particular, the renormalon-based model
became popular. Through the resummation of the factorially divergent perturbative
series for the twist-two coefficient functions, infrared renormalons generate an
{\em ambiguous} higher-twist contribution. The renormalon approach to power
corrections~\cite{Beneke} is based on the consistency of the theory as well as
that of the OPE: all ambiguities in the resummed perturbative calculation must
eventually cancel out. Therefore, the presence of ambiguities $\sim
\mu^{2n}/Q^{2n}$ in the twist-two coefficient functions implies the existence of
genuine power corrections of the same form, that is of the twist$-(2n+2)$. In
this way infrared renormalons in the twist-two coefficient functions detect some
part of the higher twist contribution.

The main {\em assumption} made by the renormalon
model~(e.g.~\cite{Stein:1996wk}--\cite{Maul:1997rz}) is that the power corrections
accessible to renormalons dominate. A physical argument supporting this
assumption was never provided. A~further assumption which, in our view, makes
this model inapplicable at large $x$, is that multi-gluon contributions are
discarded by considering the large-$\beta_0$ limit~\cite{Beneke}: in this
approximation one resums a set of diagrams corresponding to a {\em single}
dressed gluon. Since at large $x$ multiple gluon emission is crucial at the
perturbative level~\re{Sudak}, it is unlikely that power corrections would be
entirely associated with a single emission. Therefore, in contrast with the above
mentioned renormalon model, we shall not base the parametrization of power
corrections at large $x$ on the single dressed gluon. In our approach multiple
gluon emission plays a major role also at the non-perturbative level.

The first part of this paper is devoted to show how the renormalon ambiguity
cancels out within the OPE, leading to unambiguous predictions to the structure
functions. The conjectured cancellation of infrared renormalons in physical
observables is fundamental to our understanding of power corrections in QCD (see
e.g.~\cite{Mueller:pa}). Nevertheless, not much has been done to test it. In this
context the example of deep inelastic scattering plays an important role: since
the DIS structure functions admit an OPE, the mechanism of cancellation can be
checked by explicit calculations. The case of the longitudinal structure function
$F_L(x)=F_1(x)-F_2(x)$ was analyzed in this spirit in~Ref.~\cite{Beneke,UnPub}.
It was shown that the renormalon ambiguity in the twist-two coefficient function
indeed cancels against that of the quadratic ultraviolet divergence of the
twist-four matrix elements. The cancellation of renormalon ambiguities for the
$F_2(x)$ structure function\footnote{The ambiguity cancellation mechanism is
believed to be general. In particular, it should apply also to the
parity-violating structure function $F_3$ as well as the spin structure
functions. Note, however, that explicit expressions for the twist four
contribution, equivalent to~\cite{Balitsky:1989bk}, are not yet available in
these cases.} by the same mechanism will be demonstrated here.

The OPE is based on the separation of short and long-distance effects into the
coefficient functions and hadronic matrix elements, respectively, by introducing
the factorization scale $\mu$. This scale provides an infrared cut-off for momenta of
particles in $\widetilde{C}_{n,a}^{i}(N,\alpha_s)$ and, at the same time, it sets up the ultraviolet
renormalization scale for $\langle O_i^{(n)}(N)\rangle$. Although both the coefficient functions and the matrix elements have
a nontrivial dependence on $\mu$, this dependence cancels in the expression for
the structure functions~\re{moments-def}. In particular, varying $\mu$ one finds
that the twist-four operators mix among themselves at the level of logarithmic ultraviolet divergences, $\sim\ln\mu^2$, as well as with the operators of twist two at the
level of quadratic ultraviolet divergences, $\sim
\mu^2$. The former divergences give rise to DGLAP evolution equations, while the
latter introduce an ambiguity in separating the contribution of twist two and
twist four. This ambiguity is compensated in \re{moments-def} by $\sim\mu^2/Q^2$
contribution to the twist-two coefficient functions induced by infrared renormalons.
Thus, the cancellation of renormalon ambiguities follows from the independence of the
DIS structure functions of the factorization scale.

Let us now return to the renormalon model and address the meaning of the
underlying assumption, that is the dominance of those power corrections which are
associated with infrared renormalons. The predictive power of the model relies on
the assertion that the dependence of the renormalon ambiguity on the Bjorken
variable $x$, or equivalently on the moment $N$, represents that of the full
higher-twist contribution. This is realized if the contribution of the
higher-twist matrix elements is dominated by the part that has the highest
possible ultraviolet divergence, and thus mixes under renormalization with the
leading twist. When viewed in the context of the OPE, the ``renormalon
dominance'' assumption is therefore understood as~\cite{UV_dom,BBM} ``ultraviolet
dominance''. It was used in~\cite{Stein:1996wk}--\cite{Akhoury:1997rt},
\cite{DMW} and \cite{Simula}--\cite{LEKN} to develop the phenomenology of power
correction to the DIS structure functions.

To leading order in the large-$\beta_0$ limit, i.e. at the level of a single dressed gluon, the ambiguities in the twist-two coefficient functions of both
$F_L(x)$ and $F_2(x)$ appear as power corrections $\sim\mu^2/Q^2$
and $\sim\mu^4/Q^4$ only. In the OPE, they are compensated by the contribution of the operators of twists four and six, respectively, that are given by the convolution of a {\em specific function} of $x$ with the {\em twist two} parton distributions. Then, the ``renormalon dominance''
assumption implies that
\begin{itemize}
\item Contributions to the structure functions of twist higher
than six are suppressed, and can be ignored.
\item The dominant twist-four and twist-six contributions are proportional to
the twist-two distribution. Therefore, these contributions can be parameterized
by a single non-perturbative parameter at each twist which controls the overall
normalization.
\end{itemize}

Because of the small number of parameters, the renormalon-based model for power
corrections seems to be very appealing phenomenologically. It is important,
however, to keep in mind the strong assumptions involved. A priori, from the OPE
point-of-view, ultraviolet dominance seems hard to justify. Higher-twist matrix elements measure multi-parton correlations within the hadronic target. They encode information on the
detailed structure of the hadron, which is inaccessible at twist two. The renormalon model reduces this rich structure to an overall normalization of the twist-two matrix elements.

The main shortcoming of the renormalon model, which is particularly
relevant at large~$x$, can be understood through perturbation theory. The renormalon ambiguity described above emerges from a specific set of diagrams in which a {\em single} dressed gluon is emitted. These diagrams dominate the large-order behaviour of the series.
However, it is well understood that at large $x$ multiple gluon emission is
important: due to the phase-space constraint, i.e. the small invariant mass of
the hadronic system, \hbox{$W^2\ll Q^2$}, the emission of soft and collinear gluons is
enhanced. In perturbation theory this results in the appearance of the Sudakov
logarithms in the coefficient functions \re{Sudak}.  These logarithms exponentiate
in the Mellin transform (moment) space owing to the factorization property of
soft and collinear gluons~\cite{Sterman}--\cite{KM}. This implies that the large-$N$ asymptotic behaviour of the perturbative coefficients is controlled by multiple emission from the primary hard parton(s). An entirely different set of diagrams than the single dressed gluon.

Analyzing the properties of the resummed perturbative expressions, one finds that
multiple soft gluon emission also affects the power
corrections~\cite{KS_DY}--\cite{KS_thrust}. The systematic way see this is by considering
the large-order behaviour of the Sudakov
exponent~\cite{Thrust_distribution}--\cite{DGE}. It turns out that the
coefficients of sub-leading Sudakov logarithms are enhanced factorially due to
infrared renormalons. This implies that any Sudakov resummation with a fixed
logarithmic accuracy is insufficient unless $W^2\gg\Lambda^2$. In the large~$x$
limit one is obliged to resum {\em all} the logarithms that are factorially
enhanced. Here infrared renormalons hit again: the resummed exponent is defined
only up to a power-suppressed ambiguity. The ambiguity must be compensated by
genuine power corrections, which {\em exponentiate} together with resummed
perturbative expansion.

The emerging structure of power corrections becomes quite different from the
renormalon model: due to multiple emission all powers of
\hbox{$1/W^{2}\sim N/Q^2$} are present
\be
\int_0^1 dx\,x^{N-2}\,F_a(x,Q^2) =
\,\widetilde{C}_{2,a}^{q}(N,\alpha_s)\,\langle O_q^{(2)}(N)\rangle\,\left[1 \,+\, \sum_{n\ge 1}
\kappa_{n}^{(a)}\, \lr{\frac{N\Lambda^2}{Q^2}}^{n}\right],
\label{moments-imp}
\ee
where $a=2,L$ and $Q^2/N$ is fixed as $N\to\infty$, and terms which are
suppressed by $1/N$ are neglected. Here, $\langle O_q^{(2)}(N)\rangle$ and
$\widetilde{C}_{2,a}^{q}(N,\alpha_s)$ are the {\em twist-two} quark-operator
matrix elements (the gluon contribution is negligible) and the corresponding
Sudakov- and renormalon-resummed coefficient functions, respectively, and
$\kappa_{n}^{(a)}\Lambda^{2n}$ are $N$-independent non-perturbative scales
related to hadronic matrix elements of twist-$2(n+1)$ operators producing the
dominant contribution as $x\to 1$.

Eventually, in order to describe the behaviour of the structure function
\re{moments-imp} at large~$x$, one has to resum an infinite set of power
corrections on the scale $Q^2/N$. In the shape function
approach~\cite{Korchemsky:1998ev,KS_thrust} the infinite set of scales
$\kappa_{n}^{(a)}\Lambda^2$, with $n=1,2,\ldots$, is replaced by a single
non-perturbative function. The series on the r.h.s.\ of~\re{moments-imp} {\em
defines} the Laplace transform of the shape function (see Eq.~\re{J-new} below),
which is a new non-perturbative distribution which governs the shape of the
structure function at large $x$. The shape function approach has already proven
useful in several processes, including Drell-Yan production~\cite{KS_DY,DGE}, and
event-shape distributions~\hbox{\cite{Korchemsky:1998ev}--\cite{Korchemsky:2000kp}}.

Independently of the renormalon analysis, we examine in this paper the structure
and the physical origin of the dominant power corrections at large $x$ within the
framework of the OPE. At first sight, the increasing number of local matrix
elements of higher-twist operators in the large-$N$ limit seems to be an obstacle
to any use of the OPE to study this limit. Nevertheless, analyzing the twist-four
contribution at leading order using its representation in terms of non-local
light-cone operators~\cite{JS}--\cite{Balitsky:1989bk} we find that, as at the
leading twist, a significant simplification occurs in this limit.

Physically, this simplification occurs owing to the dominance of particular configurations in the final and initial states. The dominating final state turns out to be the {\em same} as at twist two. The initial state differs, of course: the quark--antiquark light-cone operator of twist two is replaced by a quark--gluon--antiquark operator at twist four. However, in the dominating configuration the gluon has a small longitudinal momentum fraction. Consequently, the corresponding twist-four matrix elements essentially depend only on the light-cone separation between the quark and antiquark, just as the twist-two matrix element.

We find that the relevant correlation measured by twist four at large~$x$
corresponds to very specific configurations, which resemble twist two. This leads
us to finally to the our main conjecture, that {\em the simplification of higher
twist at large~$x$ amounts to ultraviolet dominance}, namely the dominance of
these operator matrix elements that mix under renormalization with the leading
twist. Note that, in contrast with the renormalon model, we do not assume here
the dominance of single gluon contributions; multi-gluon contributions are
relevant. The ultraviolet dominance assumption implies that the twist two matrix
element factorizes out of the twist expansion at large $x$. Moreover, for the
twist four contribution to be totaly free of factorization-scale dependence,
the resummed coefficient function must have the same logarithmic dependence on the factorization scale as the twist two coefficient function. Thus both the parton distribution function and the resummed coefficient function factorize, as summarized in~\re{moments-imp}.

This factorization implies, in particular, that the large $N$ limit has a
remarkable feature: a subset of matrix elements which dominate the twist four
contribution decouples from the rest and satisfies a closed set of evolution
equations. In fact, such an approximation is known for the twist-three
contribution to the spin structure functions of deep inelastic
scattering~\cite{Braun:2001qx}, and we plan to extend it to twist four in a
forthcoming publication.

We proceed as follows: in Section~2 we recall the standard formulation of the
OPE, and then present the known results concerning renormalon ambiguities in the
twist-two coefficient functions. In Section~3 we follow~\cite{Balitsky:1989bk}
describing twist four in terms of non-local light-cone operators and calculate
their coefficient functions in moment space. We also examine   the large-$x$
limit and show that the asymptotic behaviour of the twist four contribution in
this limit is associated with specific final states. In Section 4 we verify
explicitly the cancellation~\cite{Beneke,UnPub} between~$\sim 1/Q^2$ infrared renormalon
ambiguities at twist two and the ambiguity in the definition of twist four matrix
elements for both~$F_L$ and~$F_2$, to leading order in the flavour expansion. In
Section 5 we return to the large~$x$ limit and show that the physical picture
emerging from general considerations at twist four is consistent with the
ultraviolet dominance assumption. Our conclusions are summarised in Section 6.

\section{Twist-two analysis and infrared renormalons}

The spin-averaged structure functions, $F_L(x)$ and $F_2(x)$, of deep inelastic
scattering of a virtual photon with momentum $q$ from a hadron with momentum $p$
are defined in terms of the hadronic tensor
\be
W_{\mu\nu}(x,q^2)=\lr{g_{\mu\nu}-\frac{q_\mu q_\nu}{q^2}}
\frac{F_L(x)}{2x} -\lr{g_{\mu\nu}-\frac{p_\mu q_\nu+p_\nu
q_\mu}{(pq)} +\frac{q^2}{(pq)^2}p_\mu p_\nu} \frac{F_2(x)}{2x},
\label{structure_functions}
\ee
with $x\equiv Q^2/2pq$ and $Q^2=-q^2>0$. The latter is expressed through the
imaginary part of the Fourier transform of the time-ordered ($T$) product of two
electromagnetic currents, symmetrized over the Lorentz indices:
\ba
\label{W}
W_{\mu\nu}(x,q^2)&=&\frac {8}{\pi}\,\Im\left\{ \int d^4 y\,\e^{i\,qy} \vev{p\,|
iT\left[j_{\{\mu}(y)j_{\nu\}}(-y)\right]|p}\right\}
\ea
with $j_\mu(x)=\sum_{q=1}^{N_f} e_q \overline{\Psi}_q(x)\gamma_\mu \Psi_q(x)$ and
$N_f$ being the
number of quark flavours\footnote{In what follows we shall assume for simplicity that quarks of different
flavour have identical electric charges $e_q$.}.
 At large $Q^2$ the dominant contribution to \re{W} comes
from the vicinity of the light-cone $y^2\sim 1/Q^2 \longrightarrow  0$. This allows one to expand the
$T$-product into
the series of local composite operators with increasing twist ($\equiv$ dimension
$-$ spin), and, therefore, decreasing singularity for small $y^2$. A more efficient
(although equivalent) way of writing the OPE is based on expanding the
$T$-product in terms of non-local light-cone operators of increasing twist~\cite{Balitsky:1989bk}.
These operators serve as generating functionals for the infinite tower of high-twist
local operators. This formulation of the OPE proves to be more convenient for our
purpose.

To  leading order in the coupling constant, the twist-two contribution to the
OPE of the symmetric part of the $T$-product in \re{W} is given for $y^2\to 0$ by~\cite{Balitsky:1989bk}
\ba
\label{T_tw2}
iT\left[j_{\{\mu}(y)j_{\nu\}}(-y)\right]^{\rm tw-2}&=& -\frac{1}{16\pi^2y^4}\,
\left(y_{\mu}\partial_{\nu}+y_{\nu}\partial_\mu -g_{\mu\nu}(y_\alpha\partial_\alpha)\right)
\nonumber
\\
&\times&
\int_0^1du\,\left[1+\frac14\ln u\, y^2\partial_\beta^2+\CO(y^4)\right]
\overline{\Psi}(uy)\ysl \Psi(-uy)\,,
\ea
where $\partial\equiv\partial/\partial y$ and the sum over $N_f$ massless
flavours is assumed.
A path-ordered exponential between $-y$ and $y$ is required to make this operator
gauge invariant. We do not write it explicitly here.

Going over to hadronic matrix elements, we introduce the standard twist-two quark
distribution $q(\xi)$ by the relation
\be
\vev{p|\overline{\Psi}(y) \ysl \Psi(-y)|p}_{\mu}
=2(py)\,\int_{-1}^{1} d\xi \,\e^{2i\xi\, (py)} q(\xi,\mu^2)\,,
\label{tw2_mat_el}
\ee
where $y^2=0$ and $\mu$ defines the normalization scale of the non-local
light-cone operator. Inserting \re{T_tw2} into \re{W} and calculating the
hadronic tensor, we arrive at the well-known expressions for the Born
approximation to the twist-two part of the structure functions
\be
F_L^{(0)}(x)=0\,,\qquad F_2^{(0)}(x)= x\sum_q e_q^2 \left[q(x)+\bar
q(x)\right]\,,
\label{Born}
\ee
where $\bar q(x)=q(-x)$ stands for the antiquark distribution. Going beyond the
leading order in \re{T_tw2}, one finds that the coefficient function of the
non-local quark operator is modified by ${\cal O}(\alpha_s)$ corrections and,
moreover, the $T$-product receives contribution from non-local light-cone gluon
operators. The general expression for the twist-two contribution to the
structure functions is then given by
\be
\frac{1}{x} F_a^{\rm
tw-2}(x,Q^2)=\int_x^1\frac{d\xi}{\xi}\,C_{2,a}^q(\xi,\alpha_s(Q^2),Q^2/\mu^2)
\,q(x/\xi,\mu)+ \ldots
\label{F-2-gen}
\ee
where $a=L,2$ stand for $F_L$ and $F_2$, respectively, and $C_{2,a}^q(\xi,\alpha_s(Q^2),Q^2/\mu^2)$
are the corresponding quark coefficient functions. The ellipses stand for the antiquark and the gluon contributions.

For our purpose, the gluon contribution to \re{F-2-gen} can be safely ignored
for the following reasons. Our task in Section~4 is to verify the
cancellation of infrared renormalons at twist two and ultraviolet contribution to
the twist-four matrix elements, to the {\em leading order} of the flavour
expansion. However, as was argued in~\cite{Beneke}, renormalons appear in the
gluon matrix elements only at the next-to-leading order in the flavour expansion.
Secondly, when examining the asymptotics of the structure function $F_2(x)$ in~\re{F-2-gen} for~$x\to 1$, one finds~\cite{DDKS,SanchezGuillen:iq,Zijlstra:1992qd,Sotiropoulos:1999hy} that the gluon coefficient function is subleading as compared with that of the quark.
For example, to order $\alpha_s$, the $F_2$ coefficient functions are
\ba
\label{C22_qg}
C_{2,2}^{q, {\rm
NS}}(\xi,\alpha_s(Q^2),1)&=&\delta(1-\xi)\,+\,C_F\,\frac{\alpha_s}{\pi}\,\left[\left(\frac{\ln(1-\xi)}{1-\xi}\right)_{+}-\frac34\left(\frac{1}{1-\xi}\right)_{+}\right.
\\ \nonumber &&\left.-\frac12(1+\xi)\ln (1-\xi)-\frac12
\frac{1+\xi^2}{1-\xi}\ln\xi+\frac32+\xi-\delta(1-\xi)\left(\frac{\pi^2}{6}
+\frac94\right)\right]\\ \nonumber
C_{2,2}^{g}(\xi,\alpha_s(Q^2),1)&=&N_f\,\frac{\alpha_s}{\pi}\,\left[\frac12
\xi(1-2\xi+2\xi^2)
\left[\ln(1-\xi)-\ln(\xi)\right]-\frac12+4\xi (1-\xi)\right],
\ea
and the $F_L$ coefficient functions are
\ba
\label{C2L_qg}
C_{2,L}^{q, {\rm NS}}(\xi,\alpha_s(Q^2),1)&=&C_F\,\frac{\alpha_s}{\pi}\,\xi \nonumber\\
C_{2,L}^{g}(\xi,\alpha_s(Q^2),1)&=&N_f\,\frac{\alpha_s}{\pi}\,2\xi(1-\xi),
\ea
where the factorization scale was set as $\mu^2=Q^2$.
In both cases there is an additional suppression factor of $(1-\xi)$ for the gluon coefficient function  compared with that of the quark, and  therefore the gluon contribution can be ignored provided that the gluon density is not anomalously large compared with the quark density for $x\to 1$. We shall assume that this is
indeed the case. Phenomenological studies~\cite{MRST} shows that the gluon density decreases faster than that of the quarks, consistently with this assumption.

To see intuitively why the hierarchy between the quark and gluon coefficient function at large~$x$ holds, one has to compare the final states described by
the two contributions. For the quark contribution to $F_2(x)$ the
final state consists of the outgoing energetic quark jet of small invariant mass
$Q^2(1-x)$ surrounded by soft gluons. For the gluon contribution, coming
necessarily through the box diagram, the cloud of soft particles is bound to
contain a soft quark. Invoking the standard power-counting arguments, one finds
that the contribution of a soft quark is suppressed by a power of the energy
$\sim (1-x)$ compared with that of a soft gluon.

In addition one finds that owing to the different
quantum numbers, the $F_L$ coefficient functions are suppressed at large~$x$ by an overall
factor of $(1-x)$, compared with $F_2$. Taking the moments of \re{F-2-gen}
\be
\widetilde{F}_{a}(N,Q^2)\equiv \int_0^1 F_a(x,Q^2)\, x^{N-2} dx
= \widetilde{C}_{2,a}^{q}(N,\alpha_s) \,q_N
\label{tw2_mom}
\ee
with $q_N \equiv \int_0^1 q(x) \, x^{N-1} dx$ and $\widetilde{C}_2(N)
\equiv \int_0^1 C_2(x) \, x^{N-1} dx$, one concludes that
\be
\widetilde{F}_{L}^{\rm tw-2}(N)/\widetilde{F}_{2}^{\rm tw-2}(N)
\sim \widetilde{C}_{2,L}(N)/\widetilde{C}_{2,2}(N) ={\cal
O}(1/N)\,,
\label{hierarchy}
\ee
The validity of the hierarchy~\re{hierarchy} proves to be independent of the twist.
In particular, as we shall show below, the same hierarchy \re{hierarchy}
holds for the twist-four contribution, and it becomes a general property of the structure functions in the semi-inclusive region.

The twist-two coefficient functions $C_{2,a}^i(\xi,\alpha_s(Q^2),Q^2/\mu^2)$ have
a well-defined perturbative expansion. To any order in the coupling constant,
their perturbative expressions contain terms involving logarithmic dependence on
the factorization scale $\mu$ as well as the terms regular as $\mu\to 0$. The
former are controlled by the DGLAP evolution equations, while the latter become
sensitive to the behaviour of the coupling constant at low scale and increase
factorially at high orders in $\alpha_s$ (renormalons). Infrared renormalons make
the perturbative series Borel non-summable and thus induce ambiguous
contributions to the coefficient functions suppressed by powers of $1/Q^2$.

The logarithmic dependence of the twist-two parton distributions $q(x,\mu)$ on
the factorization scale $\mu$ reflects the $\ln\mu$ renormalization scale
dependence of the corresponding operators \re{tw2_mat_el}. Through
renormalization, different operators mix with each other. Similarly, higher twist
$n$ (with $n=4,6,\ldots$) distributions admit power-like dependence on the
factorization scale, $\sim\mu^{2n}$. This reflects the fact that the higher-twist
operators mix under renormalization with the leading twist. As a consequence, the
separation of higher-twist matrix elements from those of the leading twist
depends on the factorization scale. Obviously, the structure functions
$F_a(x,Q^2)$ should not depend on this scale. Within the OPE \re{moments-def},
the logarithmic $\mu$ dependence cancels out between the coefficient functions
and the matrix elements of the same twist, whereas the power-like $\mu^{2n}$
dependence cancels between the contributions of {\em different} twists. From this
we draw two important conclusions. The first is that in order to verify the
cancellation of infrared renormalon ambiguities in the twist-two coefficient
functions, one has to calculate the most divergent ultraviolet contribution to
the matrix elements of higher-twist operators -- the quadratic divergence for
twist four, the quartic divergence for twist six, etc. The second conclusion has
considerable implications on phenomenological applications: in general,
higher-twist contributions cannot be consistently included\footnote{A similar
conclusion was reached in~\cite{Average_thrust}.} without resumming the
perturbative expansion (which is, by itself, ambiguous). The ambiguity cancels
only when resummation is performed {\em and} power corrections are included.

\begin{figure}[htb]
\begin{center}
\centerline{{\epsfxsize15.0cm\epsfbox{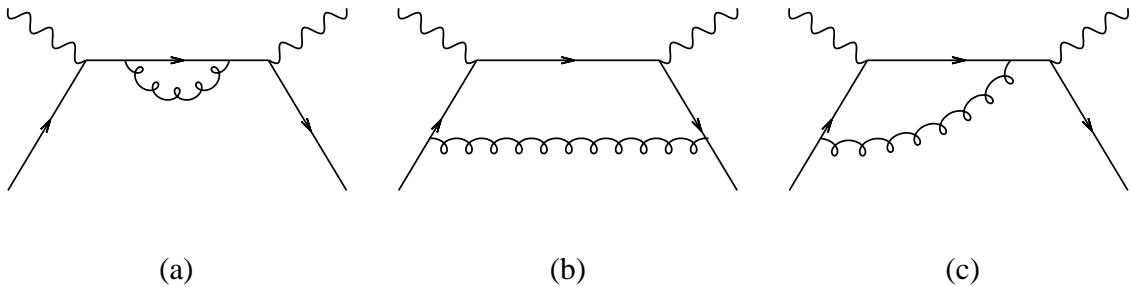}}}
\end{center}
\caption{Diagrams that contribute to the twist-two coefficient function at
leading order.}
\label{Diagrams}
\end{figure}

Renormalon contributions to the twist-two coefficient functions, $\sim
\alpha_s^{n+1}\beta_0^n\,n!$, can be extracted from the (unphysical, but technically
useful) large $N_f$ limit, since $\beta_0=(11 C_A-2N_f)/12$ is linear in $N_f$. The
leading order in the flavour expansion corresponds to the
lowest-order diagrams in figure~\ref{Diagrams}, where the internal
gluon line is dressed by any number of fermion loops.

To regularize a divergent renormalon contribution, we shall apply the Borel
regularization~\cite{Beneke}. It
amounts to representing the large-$N_f$ dressed-gluon propagator by
\[
\frac {1}{k^2}\,\frac{1}{\beta_0}\int_0^{\infty} \,du \, \exp\left(-u \ln \frac{k^2}{\Lambda^2}\right)
\]
and identifying singularities of the resulting momentum integral in the
$u$-plane located at real positive $u$. The main advantage of the Borel method is that it allows one to regularize simultaneously the renormalon contribution to the coefficient functions {\it and\/} power-like ultraviolet divergences to the higher-twist matrix elements.

The renormalon contribution to the twist-two coefficient functions, $C_{2,2}^{q}$
and $C_{2,L}^{q}$, has been calculated in the single dressed gluon approximation
in Refs.~\cite{Stein:1996wk}--\cite{Maul:1997rz} (see also \cite{Akhoury:1997rt} and~\cite{DMW}). It generates power-suppressed corrections to
the coefficient functions of the form $\sim 1/Q^2$ and $\sim 1/Q^4$ only and,
therefore, produce ambiguous contribution to the structure functions of twist
four and twist six, respectively. In particular, the twist-four renormalon
contribution has the following form
\ba
\label{Ren_amb_FL}
\int_0^1dx\, x^{N-1}\Delta_{_{\rm IR}}\left[\frac{F_2}{2x}\right]_{\rm tw-4} \! &=& \! \frac{ C_F
}{4\beta_0}
\left[4\psi(N+1)+4\gamma+\frac{2}{1+N}+\frac{12}{2+N}-8-N\right]q_N\frac{\Lambda^2\delta}{Q^2}\,,
\nonumber
\\
\int_0^1dx\, x^{N-1}\Delta_{_{\rm IR}}\left[\frac{F_L}{2x}\right]_{\rm tw-4}  \! &=& \! \frac{ C_F
}{4\beta_0}\left[-\frac{4N}{N+2}
\right]
\,q_N\frac{\Lambda^2\delta}{Q^2}\,,
\ea
with $\psi(x)=d\ln\Gamma(x)/dx$ and $\gamma$ being the Euler constant. Here, the
notation was introduced for the dimensionless constant $\delta$, defined by the integral
$\delta\equiv
\int_{\cal C}\!{du}/(1-u)$, which represents the difference between different
deformations of the original Borel-integration contour ($0<u<\infty$) in the
vicinity of the pole at $u=1$.

The expressions \re{Ren_amb_FL} simplify significantly at large $N$,
\ba
\label{Ren_amb_asym}
\int_0^1dx\, x^{N-1}\Delta_{_{\rm IR}}\left[\frac{F_2}{2x}\right]_{\rm tw-4} \! &=& \! - \frac{ C_F
}{4\beta_0} q_N\frac{N\Lambda^2\delta}{Q^2}\,,
\nonumber
\\
\int_0^1dx\, x^{N-1}\Delta_{_{\rm IR}}\left[\frac{F_L}{2x}\right]_{\rm tw-4}  \! &=& \! -\frac{ C_F
}{\beta_0}
\,q_N\frac{\Lambda^2\delta}{Q^2}\,,
\ea
so that the renormalon contribution to $F_L(x,Q^2)$ is suppressed at large $N$ by
a power of $N$ as compared with $F_2(x,Q^2)$. We recall that a similar property
holds for the twist-two contribution, Eq.~\re{hierarchy}. In addition, Eq.~\re{Ren_amb_asym} suggests that, in agreement with our expectations \re{moments-imp}, the effective parameter of the twist expansion in the limit $x\to 1$ is given by $N/Q^2$ rather than $1/Q^2$.
An expression similar to~\re{Ren_amb_FL} is available for the twist-six renormalon ambiguity \cite{DasW_DIS}. In the Borel plane, this ambiguity appears as an additional pole, at $u=2$.
At large $N$, this contribution contains an additional factor of $N/Q^2$ with respect to twist four. Again, this holds for both $F_2(x,Q^2)$ and $F_L(x,Q^2)$.

Consistency of the OPE of the structure functions requires that
renormalon ambiguities \re{Ren_amb_FL} will cancel in \re{moments-def} against
the ultraviolet divergent contribution to the twist-four matrix elements. For the longitudinal
structure function, $F_L(x,Q^2)$, this cancellation was demonstrated
in~\cite{Beneke,UnPub}.
Below we will conduct a similar analysis for case of the
structure function $F_2(x,Q^2)$, which according to \re{Ren_amb_asym} is expected
to have a more singular behaviour for $x\to 1$.
The cancellation of the twist-six renormalon ambiguities with ultraviolet divergent
contribution to the twist-six matrix elements requires a complete analysis of
twist-six operators and their coefficient functions. It is therefore more
involved and will not be discussed here.

\section{Twist-four analysis}

To demonstrate the cancellation of the renormalon ambiguities \re{Ren_amb_FL}, we
will need the expression for the twist-four part of the structure functions. This
relies on the twist-four contribution to the symmetric part of the time-ordered
product~(\ref{T_tw2}). To  leading order in the coupling constant it is given
by~\cite{JS}--\cite{Balitsky:1989bk},
\be
iT\left[j_{\{\mu}(y)j_{\nu\}}(-y)\right]^{\rm tw-4}=\frac{1}{128\pi^2}\left({\cal
T}_{2,\,\mu\nu}+ {\cal T}_{L,\,\mu\nu}+ \cdots\right)
\label{T-prod1}
\ee
with two terms ${\cal T}_{L,\,\mu\nu}$ and ${\cal T}_{2,\,\mu\nu}$
representing the separate projections of the longitudinal and the transverse
components, respectively, and the ellipses denoting the operators
containing total
derivatives which therefore having vanishing forward matrix elements. The operators
${\cal T}_{a,\,\mu\nu}$ (with $a=2,\,L$) are defined for $y^2\to 0$ as
\be
{\cal T}_{L,\,\mu\nu}=\int_0^1du\left[\frac{4g_{\mu\nu}}{y^2}u(1+\ln u)\, Q_1(uy)
+u\ln u\, \partial_\mu\partial_\nu \lr{\ln{y^2}\, Q_1(uy)}\right]
\label{T_L}
\ee
and
\ba
\label{T_2}
{\cal T}_{2,\,\mu\nu}&=&\int_0^1du\bigg\{-\ln
y^2\,\partial_\mu\partial_\nu\left[u\ln u \,Q_1(uy)+
\frac{1-u}{u^2} \,Q_2(uy)\right]
\nonumber
\\
&&+\frac1{y^2}\lr{y_\mu\partial_\nu+y_\nu\partial_\mu-g_{\mu\nu} (y\cdot
\partial)}\left[-u(1+3\ln u) \,Q_1(uy)+\frac1{u} \,Q_2(uy)\right]\bigg\}\,.
\ea
Here, $Q_1(y)$ and $Q_2(y)$ are two specific combinations of non-local twist-four
light-cone operators, which will be given later on.

Inserting \re{T_L}, \re{T_2} and \re{T-prod1} into the expression for the
hadronic tensor, \re{W}, we parameterize, in analogy with \re{tw2_mat_el}, the
matrix elements of the operators $Q_1(y)$ and $Q_2(y)$ in terms of the twist-four
distributions $D_1(\xi)$ and $D_2(\xi)$, respectively,
\be
\vev{p| Q_i(y)|p}
=i \,\int_{-1}^{1} d\xi \,\e^{2i\xi\, py} D_i(\xi),
\label{tw4_mat_el}
\ee
with $y^2=0$. The symmetry of the $T$-product, Eq.~\re{T-prod1}, under $y\to -y$
implies that $Q_i(-y)=Q_i(y)$, and, as a consequence, $D_i(\xi)=D_i(-\xi)$ (with
$i=1,2$). The functions $D_i(\xi)$ defined is this way describe the momentum
fraction distribution inside the hadron and vanish outside the interval $-1<
\xi<1$. However, in contrast to the twist-two distributions, they
do not have probabilistic interpretation and, in general, are not positive
definite. As we will show below, they can be expanded into a sum over
multi-parton configurations with definite weights, given by the twist-four
coefficient functions.

Calculating the twist-four contribution to the hadronic tensor according to
(\ref{W}), we perform a Fourier transform of the forward matrix element of
\re{T-prod1} and take its imaginary part. Matching the resulting expression into
the definition of the structure functions, Eq.~\re{structure_functions}, we
obtain after some algebra the following expressions for the twist-four contributions to the structure functions
\ba
\label{F-2-x}
\left.\frac{F_L(x)}{2x}\right|_{\rm tw-4}&=&\frac{1}{4q^2}
\int_x^1\frac{d\xi}{\xi}  \,{\xi^2}
\lr{1+\ln{\xi}}\,D_1(x/\xi)\,,
\nonumber
\\
\left.\frac{F_2(x)}{2x}\right|_{\rm tw-4}&=&\frac{1}{16q^2}\left[
x\left(D_2(x)-D_1(x)\right)+\int_x^1\frac{d\xi}{\xi}\,
{\xi^2}\lr{5+6\ln{\xi}}\,D_1(x/\xi)
\right]\,.
\ea

The expressions obtained for the
twist-four contribution to the structure functions
have a striking similarity to the twist-two contribution given by~\re{F-2-gen}.
Indeed, one can rewrite~\re{F-2-x} in the same form as~\re{F-2-gen} by replacing
twist-two parton densities by their twist-four counterparts~$D_i(\xi)$ and
defining the corresponding coefficient functions~$C_{4,a}(\xi)$. This analogy
proves to be very useful as it reveals the important difference between
the two terms
entering the second relation in \re{F-2-x}: the first term where the distribution amplitude is evaluated at~$\xi=x$, and the second, which involves an integral over
the partonic momentum fraction~$\xi$. In the case
of twist two, these two terms are associated with the contribution to the
coefficient functions due to virtual and real perturbative corrections to the
partonic cross section, respectively. The former corresponds to a single-parton
final state, $C_{2,a}(\xi)\sim \delta(1-\xi)$, and it is singular at $x=1$. In
general, to high orders in the perturbation
 expansion, the singular behaviour of the
coefficient functions for $x\to 1$ is associated with particular partonic
(short-distance) final states in which the recoiling quark creates a narrow jet.
Generalization of this analysis to the twist-four contribution is
straightforward.

Following \cite{Jaffe}, we can rewrite the twist-four contribution to the
structure function as the sum over all possible final states. Then, the two terms in
the above-mentioned expression for $F_2(x)$  correspond to two different
short-distance final states: a single recoiled quark state and a state in which
this quark is accompanied by an additional gluon carrying the momentum fraction
$1-\xi$. We also notice that the coefficient functions entering the expressions
for $F_L(x)$ and the second term in the expression for $F_2(x)$ in \re{F-2-x} are
not singular as $\xi\to 1$. As in the case of
 twist two, this suggests that the
emitted gluon cannot be soft and, as a consequence, the corresponding
contribution to the structure functions is subleading at large $x$  compared
with the first term in the second relation in \re{F-2-x}.

Indeed, let us examine the twist-four contributions~\re{F-2-x} in the large
$x$ limit. It is convenient to consider the expressions~\re{F-2-x} in
moment space
\ba
\label{F-mom}
\int_0^1 dx\, x^{N-1} \left.\frac{F_L(x)}{2x}\right|_{\rm tw-4}
&=& \frac{1}{4q^2}\,\frac{N+1}{(N+2)^2}\,\widetilde{D}_1(N+1)\,,\\
\int_0^1 dx\, x^{N-1} \left.\frac{F_2(x)}{2x}\right|_{\rm tw-4}
&=& \frac{1}{16q^2}\,\left[ \widetilde{D}_2(N+1)-\frac{N(N-1)}{(N+2)^2}\,
\widetilde{D}_1(N+1)\right]\,,
\nonumber
\ea
where $\widetilde{D}_i (N) \equiv \int_0^1 d \xi\,\xi^{N-1}\, D_i(\xi)$. Assuming that
the moments $\widetilde{D}_{1}(N+1)$ and $\widetilde{D}_{2}(N+1)$
have a similar behaviour at large $N$, we
conclude from \re{F-mom} that the twist-four contribution to $F_L(x)$ is
suppressed by a power of $N$  compared with that of $F_2(x)$. We recall that
the same property holds for the twist-two contribution to the structure
functions, Eq.~\re{hierarchy}, including the renormalon contribution
\re{Ren_amb_asym}.

We also find that for large $x$, the twist-four
contribution to $F_2(x)$ (this is not so for $F_L(x)$) has a natural
approximation,
\be
\left.\frac{F_2(x)}{2x}\right|_{\rm tw-4}\simeq\frac{1}{16q^2}\,
x\left(D_2(x)-D_1(x)\right)\,.
\label{F-2-x_app}
\ee
Remarkably enough, this is similar to the expression for the
twist-two structure function to the lowest order in $\alpha_s$, Eq.~\re{Born}.
 As  explained above, the reason for this is that in both cases the
partonic {\it final\/}
 states which determine the large-$x$ asymptotics of the coefficient
functions look alike: they correspond to the  propagation of a single recoiled
quark. Calculating high order corrections to the twist-four coefficient
functions, we expect that, as in
 the twist-two situation, the large-$x$ asymptotics will correspond to the propagation in the final state of narrow
jet initiated by the recoiled quark. The difference between twists resides, at
large $x$, in the different {\it initial\/} states. For the twist two, the initial
state consists of a single parton and the coefficient function is given by the
squared modulus of the diagonal transition amplitude, or, equivalently, the
partonic cross section. For twist four, there can be one or more partons in
the initial state, so that the coefficient function is given by the sum of
diagonal transition amplitudes as well as their interference. This implies that
twist-four coefficient functions do not have an interpretation of the partonic
cross section but rather of correlations between different multi-parton states
within the hadron~\cite{Jaffe}.

\begin{figure}[htb]
\centerline{{\epsfxsize8.0cm\epsfbox{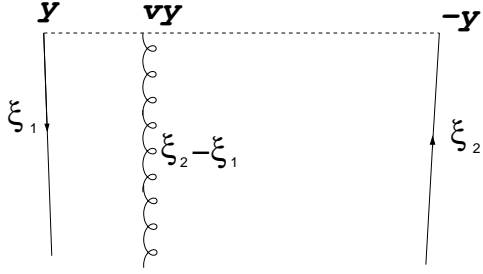}}}
\vspace*{-30mm}
\caption{Forward matrix elements of the non-local light-cone operators
${\cal O}_{3,7}(v;y)$.}
\label{qgq}
\end{figure}

Let us examine the properties of twist-four distributions $D_i(x)$ in more
detail. According to Eq.~\re{tw4_mat_el}, these distributions are defined through
hadronic matrix elements of non-local light-cone twist-four operators $Q_i(y)$.
Their expressions have been worked out in~\cite{JS}--\cite{Balitsky:1989bk} and
are given by\footnote{These expressions were taken from
Ref.~\cite{Balitsky:1989bk} and a few misprints were corrected.}
\ba
\label{A12}
\nonumber
Q_1(y)&=&\int_{-1}^1dv\,\bigg[ 4 \,{\cal O}_3(v;y)-2i(1-v^2)\,{\cal O}_7(v;y) \\
\nonumber &+& 4\, \int_{-1}^vdt(1-vt)\,{\cal O}_5(v,t;y)+(v-t)\,{\cal O}_6(v,t;y)
\bigg] \,\,+\,\, (y\leftrightarrow -y)\\ \nonumber
Q_2(y)&=&\int_{-1}^1dv\,\bigg[-4i(1-v^2)\,{\cal O}_7(v;y) +4\,\int_{-1}^vdt\,
\bigg\{
(2-v+t-2vt)\,{\cal O}_5(v,t;y)
\\
&+& (v-t)\,{\cal O}_6(v,t;y)+    {\cal O}_1^{\rm sym}(v,t;y)+ {\cal O}_2^{\rm
sym}(v,t;y)\bigg\}\bigg]\,\,+\,\, (y\leftrightarrow -y)\,.
\ea
Here, the notations for the ``canonical'' basis of light-cone operators ${\cal
O}_i$ are according to~\cite{JS}. They can be classified according to their
partonic content and are defined as follows. The quark-antiquark-gluon operators
are
\ba
\label{O37}
{\cal O}_3(v;y)&=&\frac12 g \epsilon_{\mu\nu\rho\eta}\overline{\Psi}(y) y^\mu
\gamma^\nu
\gamma_5  G^{\rho\eta}(vy) \Psi(-y)
\\
{\cal O}_7(v;y)&=&g \overline{\Psi}(y) \ysl  y_\nu D_\mu \nonumber G^{\mu\nu}(vy)
\Psi(-y)\,,
\ea
where $y^2=0$ and $G_{\mu\nu}\equiv t^a G_{\mu\nu}^a$ with $t^a$ being the color matrices in the quark representation. Other operators include a quark, an antiquark and two
gluons,
\ba
{\cal O}_5(v,t;y)&=&g^2 \overline{\Psi}(y)y_\alpha G^{\alpha\eta}(vy)
y_{\beta} G^{\beta\eta}(ty) \ysl \Psi(-y)\\ \nonumber
{\cal O}_6(v,t;y)&=&i g^2 \epsilon_{\mu\nu\rho\eta} \overline{\Psi}(y) y^{\rho}\,
y_\alpha G^{\alpha\mu}(vy) y_\beta G^{\beta\nu}(ty)\gamma^\eta\gamma_5\Psi(-y),
\ea
and $C$-even combination (sym) of two quark-antiquark pairs,
\ba
\label{sym_O12}
{\cal O}_1^{\rm sym}(v,t;y)&=&\frac{ig^2}4\!\left[\overline\Psi(y) t^a
\ysl\Psi(vy) - \overline\Psi(vy) t^a \ysl \Psi(y)\right]
\\
&&\,\,\times
\left[\overline\Psi(ty) t^a
\ysl\Psi(-y) - \overline\Psi(-y) t^a \ysl \Psi(ty)\right]
\nonumber \\
{\cal O}_2^{\rm sym}(v,t;y) &=& \frac{ig^2}4\!\left[\overline\Psi(y) t^a
\ysl\gamma_5\Psi(vy) + \overline\Psi(vy) t^a \ysl \gamma_5\Psi(y)\right]
\nonumber \\
&&\,\,\times \left[\overline\Psi(ty) t^a
\ysl\gamma_5\Psi(-y) + \overline\Psi(-y) t^a \ysl\gamma_5 \Psi(ty) \right]\,.
\nonumber
\ea
Inserting~\re{A12} into~\re{tw4_mat_el}, we can relate the distributions
$D_i(\xi)$ to the matrix elements of the basis twist-four operators ${\cal
O}_{\alpha}(v;y)$, which, in turn, can be parameterized by introducing the set of
twist-four multi-parton distributions (see figure~\ref{qgq}),
\ba
\label{T37}
\vev{p|{\cal O}_3(v;y)|p}&=&2 (py)\int d\xi_1 d\xi_2\,
\e^{i(py)[\xi_1(1-v)+\xi_2(1+v)]}\, T_3(\xi_1,\xi_2) \nonumber\\
\vev{p|{\cal O}_7(v;y)|p}&=&2 (py)^2\int d\xi_1 d\xi_2\,
\e^{i(py)[\xi_1(1-v)+\xi_2(1+v)]}\, T_7(\xi_1,\xi_2)\,.
\ea
Expanding  both sides of~\re{T37} in powers of~$y$ we find that, as in the case
of twist two, the moments of the distributions $T_i(\xi_1,\xi_2)$ are related to
the matrix elements of {\it local\/} twist-four operators. The distribution
functions corresponding to the remaining operators from the canonical basis are
defined similarly.

The distributions $T_{3,7}(\xi_1,\xi_2)$ are the analogue of the twist-two
distribution $q(\xi)$ introduced in (\ref{tw2_mat_el}). They describe the
correlations inside the hadron between
 a state containing a quark and a gluon carrying the momentum fractions $\xi_1$, $\xi_2-\xi_1$, respectively, and a state containing quark carrying momentum fraction $\xi_2$.
These distributions vanish outside the ``physical'' region
$-1<\xi_1,\,\xi_2,\,\xi_2-\xi_1 < 1$. Using the hermiticity properties of the
operators \re{O37}, $\left[{\cal O}_3(v;y)\right]^\dagger = - {\cal O}_3(-v;-y)$
and $\left[{\cal O}_7(v;y)\right]^\dagger = {\cal O}_7(-v;-y)$ we deduce from
\re{T37} that
\be
T_3^*(\xi_1,\xi_2)=T_3(\xi_2,\xi_1)\,,\qquad
T_7^*(\xi_1,\xi_2)= T_7(\xi_2,\xi_1)\,.
\label{T-real}
\ee
In addition, time-reversal invariance of the matrix elements in \re{T37} implies
that\footnote{We thank O.~Teryaev for useful discussion on this point.}
\be
T_3(\xi_2,\xi_1)=T_3(\xi_1,\xi_2)\,,\qquad T_7(\xi_2,\xi_1)=T_7(\xi_1,\xi_2)\,.
\label{T-time}
\ee
{}From \re{T-real} and \re{T-time} one deduces that the distributions
$T_{\alpha}(\xi_1,\xi_2)$ are in fact real-valued symmetric
functions~\cite{Teryaev}.

Taking into account \re{T37}, it becomes straightforward to express the
distributions $D_i(\xi)$ defined in \re{tw4_mat_el} in terms of the canonical
twist-four distributions $T_i(\xi_i)$. To this end  one substitutes \re{T37} into
\re{A12} and matches the result of $v$ integration into the r.h.s.\ of
\re{tw4_mat_el}. Expanding the both sides of the resulting relation in powers of
$(py)$, we obtain the following expression for the {\it odd\/} moments $N$ of the
distributions
\ba
\label{D-moments}
&&\int_0^1 d\xi\,\xi^{N-1}\, D_1(\xi) = -\widetilde{T}_3(N)+\widetilde{T}_7(N)+
\ldots
\nonumber
\\
&&\int_0^1 d\xi\,\xi^{N-1}\, D_2(\xi)= 0\cdot \widetilde{T}_3(N)+2 \cdot\widetilde{T}_7(N)+ \ldots,
\ea
where the ellipses stand for the contribution of the remaining distributions from the
canonical basis, which we demonstrate in the next Section to be irrelevant for the cancellation of renormalon ambiguities to leading order in the flavour expansion. Here
we have introduced the notation for the generalized moments:
\be
 \widetilde{T}_\alpha(N)= 16\int d\xi_1d\xi_2\, T_\alpha(\xi_1,\xi_2)\,
\Phi_{\alpha,N-2}(\xi_1,\xi_2),
\label{gen-mom}
\ee
with $\alpha=3,\, 7$. $\Phi_{3,N}(\xi_i)$ and $\Phi_{7,N}(\xi_i)$ are
homogenous polynomials of degree $N$ in the momentum fractions $\xi_i$. Inserting
the expressions \re{A12} for the non-local twist-four operators in terms of the
canonical operators ${\cal O}_\alpha$ into \re{tw4_mat_el} and \re{D-moments}
uniquely determines the form of the polynomials $\Phi_{\alpha,N}$:
\ba
\label{Phi37}
\Phi_{3,N}(\xi_1,\xi_2)&=&\frac{\xi_1^{N+1}-\xi_2^{N+1}}{\xi_1-\xi_2}
=\sum_{k=0}^N\xi_1^k\xi_2^{N-k}
\nonumber
\\
\Phi_{7,N}(\xi_1,\xi_2)&=&\frac1{N+2}\partial_{\xi_1}\partial_{\xi_2}
\Phi_{3,N+1}(\xi_1,\xi_2)
=\sum_{k=0}^{N-1}\frac{(k+1)(N-k)}{N+2}\xi_1^k\xi_2^{N-k-1}\,.
\ea
Since the functions $\Phi_{3,N}(\xi_i)$ and $\Phi_{7,N}(\xi_i)$ are symmetric
under interchange of $\xi_1$ and $\xi_2$, the moments \re{gen-mom} only receive
nonzero contribution  from the real part of the distributions $T_i(\xi_i)$, in
virtue of \re{T-real}.

Repeating a similar calculation, one can obtain the expressions for the coefficient
functions $\Phi_{\alpha,N}(\xi_i)$ (with $\alpha=1,2,5,6$), defining the
contribution of the remaining twist-four canonical operators to the r.h.s.\ of
\re{D-moments}.

Using \re{F-mom} and \re{D-moments} we get
\ba
\label{F-mom1}
\int_0^1 dx\, x^{N-1} \left.\frac{F_L(x)}{2x}\right|_{\rm tw-4}
&=&
\frac{1}{4q^2}\frac{N+1}{(N+2)^2}\left[-\widetilde{T}_3(N+1)+\widetilde{T}_7(N+1)+\ldots\right]\,,\\
\int_0^1 dx\, x^{N-1} \left.\frac{F_2(x)}{2x}\right|_{\rm tw-4}
&=& \frac{1}{16q^2}\frac1{(N+2)^2}
\nonumber
\\
&&\hspace*{-10mm}\times
\left[N(N-1)\, \widetilde{T}_3(N+1)
+(N+1)(N+8)\, \widetilde{T}_7(N+1)+\ldots\right]\,.
\nonumber
\ea
These are the general expressions for the contribution of the operators ${\cal
O}_3(v;y)$ and ${\cal O}_7(v;y)$ to the moments of the structure functions. The moments $\widetilde{T}_\alpha(N)$, Eq.~\re{gen-mom}, are given by
the sum over matrix elements of local twist-four operators generated by the
non-local light-cone operators, ${\cal O}_\alpha(v;y)$. Each local operator enters
into the sum with the weight given by the coefficients of the
polynomials $\Phi_{\alpha,N}(\xi_i)$, expanded in powers of momentum fractions.

\section{Cancellation of ambiguities}

The generalized moments entering the r.h.s.\ of~\re{F-mom1} depend on the
factorization scale $\mu$ that sets up an ultraviolet cut-off for the matrix
elements of the twist-four operators~${\cal O}_\alpha(v;y)$. We now proceed to
calculate the quadratically divergent,~$\sim \mu^2$, contribution to the
distributions $T_\alpha(\xi_i)$. According to~\re{F-mom1}, this induces an
ambiguity in the definition of the twist-four structure functions, which should
cancel against the infrared renormalon contribution, Eq.~\re{Ren_amb_asym}.

To match the renormalon calculation of twist two, we must calculate the
ultraviolet contribution to the twist-four distributions in the same
approximation, that is to leading order in the flavour expansion. Since this
contribution does not depend on the particular choice of the matrix element, it
suffices to consider the matrix elements of the operators $\vev{p\,|{\cal
O}_\alpha(v;y)|p}$ between {\em quark states}, with the quark being off-shell,
$p^2 < 0$, and $y^2=0$. To calculate the quadratic divergences we have to
introduce a regularization. To be consistent with the twist-two calculation of
the coefficient functions, we are obliged to use the same regularization, namely
the Borel representation of the large-$N_f$ dressed-gluon propagator. The Borel
variable $u$ regularizes the ultraviolet divergent integrals over the gluon
momenta and quadratic divergence amounts, in this regularization, to a pole at
$u=1$, leading to an ambiguity once the Borel integral is evaluated.

\begin{figure}[htb]
\vspace*{-30mm}
\begin{center}
\mbox{\kern-0.5cm\epsfig{file=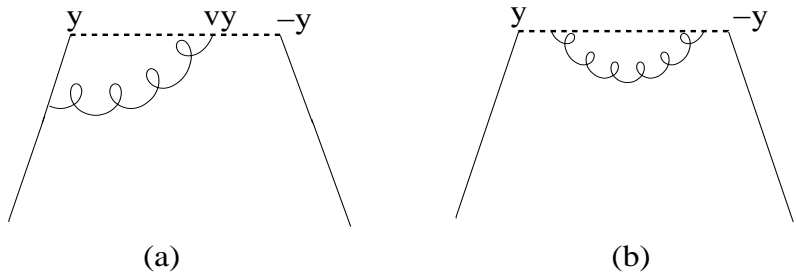,height=15truecm,width=10.5truecm,angle=
-90}}
\end{center}
\vspace*{-30mm}
\caption{Contributions to the quadratic ultraviolet divergence of twist-four
matrix elements. Insertion of fermion loops into the gluon lines is assumed.
The contribution~(b), corresponding to non-local operators with a
quark-antiquark pair and two gluons, vanishes.}
\label{twist4_diagrams}
\end{figure}

\begin{figure}[htb]
\centerline{{\epsfxsize15.0cm
\epsfbox{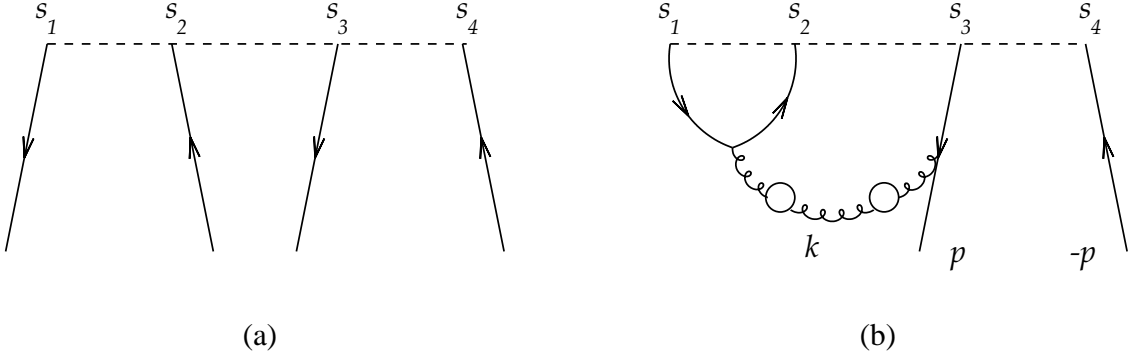}}}
\caption{(a) The non-local four-quarks operators ${\cal O}_{1,2}$;
(b) a diagram contributing to the quadratic divergence
of the matrix element of ${\cal O}_{1,2}(s_1,s_2;s_3,s_4)$ in the large-$N_f$ limit.}
\label{non_local}
\end{figure}

Examining the matrix elements of the non-local light-cone operators,
$\vev{p\,|{\cal O}_\alpha(v;y)|p}$, defined in equations~\re{O37}--\re{sym_O12}, we find
that, to leading order in the flavour expansion, only the quark--gluon--antiquark operators ${\cal
O}_3(v;y)$ and ${\cal O}_7(v;y)$ (fig.~\ref{qgq}) contribute to the quadratic divergence, and, as
a consequence, mix with the twist-two operators. The relevant diagrams are shown
in figures~\ref{twist4_diagrams} and~\ref{non_local}.

The irrelevance of the operators containing two gluons, ${\cal O}_5(v,t;y)$ and
${\cal O}_6(v,t;y)$, was pointed out in~\cite{Beneke} for the case of
$F_L(x)$. It turns out that this holds also for~$F_2(x)$. Let us briefly explain why ${\cal O}_{5,6}(v,t;y)$ are not relevant.
These operators mix with the leading
twist only if both gluons are contracted. We recall that the twist-two renormalon
contribution was calculated to leading order in the flavour expansion, so we must
apply the same approximation here. If both gluons are contracted to the quark or
antiquark lines the contribution is subleading in~$N_f$ and therefore irrelevant.
There remains the possible contribution where the gluons are contracted one to
the other, as in figure~\ref{twist4_diagrams}~(b). However, as noted
in~\cite{Beneke}, this diagram vanishes for $y^2=0$, since there is no external
scale.

The vanishing of the contribution of the four-quark operators ${\cal O}_{1,2}^{\rm sym}(v,t;y)$ of Eq.~\re{sym_O12} is less obvious.
The four quark operators (see figure~\ref{non_local}~(a)) have the form
\ba
{\cal O}_1(s_1,s_2;s_3,s_4)&=&i g^2\!\left[\overline\Psi(s_1) t^b \ysl 
\Psi(s_2)
\right]
\left[\overline\Psi(s_3) t^b \ysl 
\Psi(s_4)\right]
\nonumber\\
{\cal O}_2(s_1,s_2;s_3,s_4)&=&i g^2\!\left[\overline\Psi(s_1) t^b \ysl
\gamma_5
\Psi(s_2)\right]
\left[\overline\Psi(s_3) t^b \ysl \gamma_5
\Psi(s_4)\right].
\ea
In order for these operators to mix with the twist two operator~(\ref{T_tw2}),
one of the quarks must be contracted to one of the antiquarks. In addition, for the contribution to be relevant in the large $N_f$ limit a dressed gluon must be included.
Contracting between a quark and an antiquark of different pairs, e.g. $\Psi(s_2)$ and $\overline\Psi(s_3)$, and adding a dressed gluon with $n$ fermion loops, the contribution is ${\cal O}(g^4 (g^2 N_f)^n)$. Being sub-leading in $N_f$, this contribution is irrelevant. On the other hand, contracting between a quark and an antiquark within a pair, e.g. $\overline \Psi(s_1)$ and $\Psi(s_2)$ as shown in Fig.~\ref{non_local}~(b), an additional factor of $N_f$ is gained so the contribution is ${\cal O}(g^4 N_f (g^2 N_f)^n)$, which is relevant in the large $N_f$ limit.
Calculation of this diagram shows that~$\vev{p\,|{\cal O}_{1,2}(s_1,s_2;s_3,s_4)|p}$ is indeed quadratically divergent.
Nevertheless, because of the symmetry property
\be
\vev{p|{\cal O}_{1,2}(s_1,s_2;s_3,s_4)|p}=\theta_{1,2} \vev{p|{\cal O}_{1,2}(s_2,s_1;s_3,s_4)|p}
\ee
with $\theta_1=-\theta_2=1$, the {\em symmetrized} matrix element~$\vev{p\,|{\cal
O}_{1,2}^{\rm sym}(v,t;y)|p}$ is free of any quadratic divergence. As a
consequence the four-quark operators do not contribute to the quadratic
divergence in~\re{F-mom1}.

\begin{figure}[htb]
\centerline{{\epsfxsize13.0cm
\epsfbox{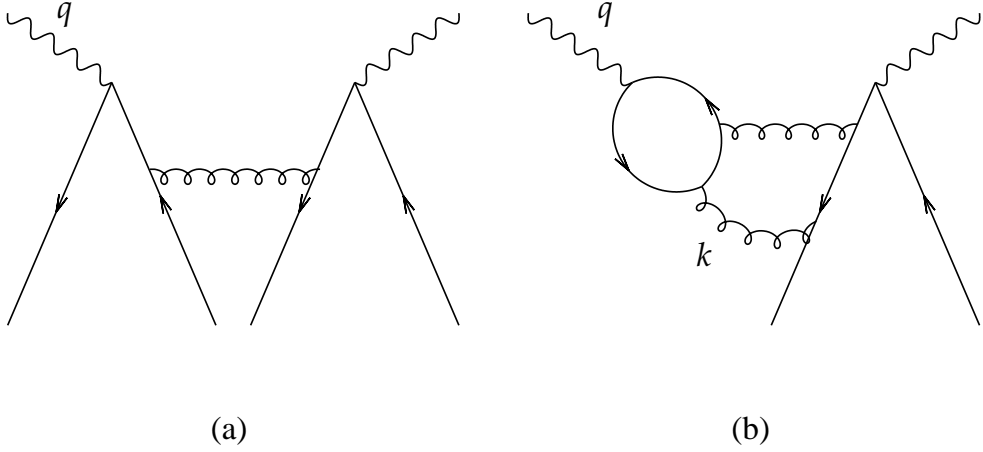}}}
\caption{(a) Local Feynman diagram from which four-quark operators emerge by taking the operator product expansion; (b) The origin of figure~\ref{non_local}~(b).}
\label{local_4quark}
\end{figure}

To understand this property, it is useful to return to the construction of the OPE of eqs.~\re{T-prod1}--(\ref{T_L})
as an expansion of local Feynman diagrams in powers of
$1/Q^2$. The four-quark operators~${\cal O}_{1,2}^{\rm sym}(v,t;y)$ emerge from
the diagram in figure~\ref{local_4quark}~(a).
The possibly relevant contributions of the non-local four-quark operators of figure~\ref{non_local}~(b) correspond,  before expansion, to the diagram in figure~\ref{local_4quark}~(b). The latter vanishes since it contains a
quark loop attached to one photon and two gluons, which is odd under (electric)
charge conjugation. The symmetrized form of~\re{sym_O12} which nullifies its contribution is thus a manifestation of charge-conjugation invariance.

We conclude that to leading order in the flavour expansion, the structure
functions~\re{F-mom1} receive a quadratically divergent contribution only from
the generalized moments $\widetilde{T}_{3,7}(N+1)$, or equivalently from the
distributions $T_{3,7}(\xi_1,\xi_2)$. The quadratically divergent contribution to
these distributions, $\Delta_{\UV} T_{3,7}(\xi_1,\xi_2)$, was calculated
in~\cite{Beneke,UnPub}.

Evaluating the diagram shown in Fig.~\ref{twist4_diagrams} (a)
with a dressed gluon and an off-shell external quark $p^2<0$ at $y^2=0$,
where the Feynman rules for the non-local operators are defined by~(\ref{O37}),
one gets
\ba
\Delta_{\UV}
T_3(\xi_1,\xi_2)&=&-\frac{C_F}{4\beta_0}\,\Lambda^2\delta\,
\left\{\frac1{\xi_1}\lr{1+\frac{\xi_2}{\xi_1}}q(\xi_1)
\theta(\xi_1-\xi_2)+(\xi_1\leftrightarrow \xi_2)\right\}
\theta(\xi_1)\theta(\xi_2)
\nonumber \\
\Delta_{\UV}
T_7(\xi_1,\xi_2)&=&2\frac{C_F}{4\beta_0}\,\Lambda^2\delta\,
\left\{\frac{\xi_2}{\xi_1}q(\xi_1)
\theta(\xi_1-\xi_2)+(\xi_1\leftrightarrow \xi_2)\right\}
\theta(\xi_1)\theta(\xi_2)\,,
\label{T-quad}
\ea
with the parameter of Borel regularization $\delta$ defined in \re{Ren_amb_FL} and
$q(\xi)$ being the twist-two quark distribution function.
Next, the ultraviolet  ambiguity of
$\widetilde{T}_{3,7}(N+1)$ can be evaluated by substituting \re{T-quad} into
\re{gen-mom} and performing the $\xi_i$ integrals with the coefficient functions
$\Phi_{3,7}(\xi_i)$ defined in \re{Phi37}. The result is
\ba
\label{a-37-UV}
\Delta_{\UV}\widetilde{T}_3(N+1)&=&\frac{C_F q_N}{\beta_0}\, \Lambda^2\delta
\left[2\psi(N+1)+2\gamma-1+\frac1{N+1}\right]\,\\ \nonumber
\Delta_{\UV}\widetilde{T}_7(N+1)&=&\frac{C_F q_N}{\beta_0}\, \Lambda^2\delta
\left[N-2\psi(N+1)-2\gamma+2-\frac2{N+1}\right]\,.
\ea

Finally, substituting \re{a-37-UV} into \re{F-mom1} we calculate the ultraviolet ambiguity of the structure functions
\ba
\label{F-fin}
\nonumber
\int_0^1 dx\, x^{N-1}
\Delta_{\UV}\!\!\left[\frac{F_2}{2x}\right]_{\rm tw-4}&=&
-\frac{C_F }{4\beta_0}
\left[4\psi(N+1)+4\gamma+\frac2{N+1}+\frac{12}{N+2}-8-N\right]\,q_N\,
\frac{\Lambda^2\delta}{Q^2}
\\
\int_0^1 dx\, x^{N-1}
\Delta_{\UV}\!\!\left[\frac{F_L}{2x}\right]_{\rm tw-4}
&=&\frac{C_F }{4\beta_0}
\frac{4N}{N+2}\,q_N\,\frac{\Lambda^2\delta}{Q^2}\,.
\ea
As expected, this ambiguity cancels exactly with the corresponding infrared
renormalon ambiguity in the twist-two coefficient functions in
Eq.~(\ref{Ren_amb_FL}). We see that when the twist-two coefficient functions are
resummed and the (ultraviolet divergent part of) the twist four is included there
remains no ambiguity at the level of $1/Q^2$ corrections in the OPE~\re{moments-def}.

We conclude that to leading order in the flavour expansion, infrared renormalons
``probe'' the contribution to the longitudinal, $F_L(x,Q^2)$, and transverse
structure functions, $F_2(x,Q^2)$, coming from only two twist-four operators,
${\cal O}_3$ and ${\cal O}_7$, defined in~\re{O37}. An additional simplification
occurs for $x\to 1$: from~\re{a-37-UV} it follows that for $N\gg 1$
\be
\Delta_{\UV}\widetilde{T}_7(N+1)/\Delta_{\UV}\widetilde{T}_3(N+1)\sim N,
\label{UV-hierarchy}
\ee
and, therefore, the contribution of the operator ${\cal O}_3$ to the
structure functions is subdominant at large $x$, as far as the cancellation of the renormalon ambiguity is concerned. As we will argue in the next Section, \re{UV-hierarchy} is not limited the ultraviolet divergent contribution of the twist-four operators, but is  a general property of these operators.
Taking into account this hierarchy, we find that
to leading order in the flavour expansion, the twist-four contribution to the structure functions for $N\to\infty$ is entirely associated with the quark-antiquark-gluon operator ${\cal O}_7$,
\ba
\label{F-mom1-asym}
\int_0^1 dx\, x^{N-1} \left.\frac{F_L(x)}{2x}\right|_{\rm tw-4}
&{=}&
\frac{1}{4q^2N}\left[\widetilde{T}_7(N+1)+{\cal O}(N^0)\right]\,,\\
\int_0^1 dx\, x^{N-1} \left.\frac{F_2(x)}{2x}\right|_{\rm tw-4}
&{=}& \frac{1}{16q^2}
\left[\widetilde{T}_7(N+1)+{\cal O}(N^0)\right]\,.
\nonumber
\ea
In addition, based on~\re{a-37-UV} we know that $\widetilde{T}_7(N+1)$
contains a piece of the form
\be
\widetilde{T}_7^{\UV}(N+1)\equiv \kappa_1^{[7]} \frac{C_F}{\beta_0}\,N \,q_N\, \Lambda^2,
\ee
where we replaced the ambiguous $\delta$ by a new $N$-independent
dimensionless parameter, $\kappa_1^{[7]}={\cal O}(1)$, and, as before, we suppressed terms that are subleading at
large $N$. Under the ultraviolet dominance assumption the remaining piece is negligible,
so
\be
\widetilde{T}_7(N+1)\simeq \widetilde{T}_7^{\UV}(N+1)\,=\,\kappa_1^{[7]} \frac{C_F}{\beta_0}\,N \,q_N\, \Lambda^2.
\label{T7UVdom}
\ee
This way the normalization of the twist-four contribution~\re{F-mom1-asym} is fixed by a single non-perturbative parameter $\kappa_1^{[7]}$, as anticipated in~\re{moments-imp}.

\section{Large-$x$ behaviour of twist four}

We saw that the ultraviolet dominance assumption leads to a great simplification
of the parameterization of the twist-four contribution at large $x$: instead of a
large number of local matrix elements, which increases $N$, this contribution is
parametrized in terms of a {\em single} parameter. But is the ultraviolet
dominance assumption justified? A clear cut answer to this question cannot be
given within the framework of perturbative QCD, since it requires calculation of
the matrix elements. However, by identifying the partonic
configurations which become dominant at large $x$, we can understand the physical
origin of the simplification that occurs at large~$x$.

We already saw, independently of the ultraviolet dominance assumption, that at
large~$x$, the twist-four contribution to $F_2$ has a natural
approximation~\re{F-2-x_app} in terms of the distributions $D_i(\xi)$, evaluated
at $\xi=x$, which is analogous to the Born-level twist-two result~\re{Born}. At
both twist two and twist four the recoiling system corresponds to a single quark.
We note that for $F_L$  at twist four, just as at twist
two~\cite{Sotiropoulos:1999hy}, the recoiling system includes a hard quark and a
hard gluon. In each case, at higher orders in perturbation theory the recoiling
system is accompanied by soft and collinear radiation. The fact that the ``bare''
recoiling system is the same for twist four and for twist two opens up the
possibility that the jet function is the same, as we shall argue below. Clearly,
this condition is not sufficient. In general, the coefficient
 functions at twist four differ from those of twist two, since the latter strictly consist
 of single-particle initial states in the amplitude and the complex-conjugate amplitude, whereas the former contain more than one parton, at least in one of them. Indeed, $D_i(\xi)$, contrary to $q(\xi)$,
describes correlations of several partons in the hadron rather than a
single-parton probability distribution.

To investigate the relevant initial states probed at large $x$, let us apply \re{F-mom1} and \re{gen-mom} to express the structure functions through the twist-four distributions $T_\alpha(\xi_1,\xi_2)$ defined in \re{T37}.
In this way, we obtain the following expression for the longitudinal
structure function in the momentum fraction representation,
\ba
\label{FL-x}
&&\hspace*{-10mm}\left.\frac{F_L(x)}{2x}\right|_{\rm tw-4}=
\frac{4}{q^2}\int_{-1}^1d\xi_1d\xi_2
\bigg\{-\widehat T_3(\xi_1,\xi_2)\frac{x^2}{(\xi_2-\xi_1)\xi_2^2}\lr{1+\ln\frac{x}{\xi_2}}
\theta(\xi_2-x)
\\
&&
+\widehat T_7(\xi_1,\xi_2)\frac{x^2}{(\xi_2-\xi_1)^3\xi_2^2}\left[
\xi_2-\xi_1+(3\xi_2-\xi_1)\ln\frac{x}{\xi_2}\right]\theta(\xi_2-x)+(\xi_1\leftrightarrow \xi_2)\bigg\}
\nonumber
\ea
and a similar expression for the transverse structure function,
\ba
\label{F2-x}
&&
\left.\frac{F_2(x)}{2x}\right|_{\rm tw-4}=
\frac{1}{q^2}\int_{-1}^1d\xi_1 d\xi_2
\bigg\{
\widehat T_3(\xi_1,\xi_2)\left[\frac{x\delta(\xi_2-x)}{\xi_2-\xi_1}
-\frac{x^2\theta(\xi_2-x)}{(\xi_2-\xi_1)\xi_2^2}\lr{5+6\ln\frac{x}{\xi_2}}
\right]
\\
&&+\widehat T_7(\xi_1,\xi_2)\left[\frac{x\delta(\xi_2-x)}{(\xi_2-\xi_1)^2}
\right.
\left. +\frac{x^2\theta(\xi_2-x)}{(\xi_2-\xi_1)^3\xi_2^2}\left(
3\xi_2-5\xi_1+6(3\xi_2-\xi_1)\ln\frac{x}{\xi_2}\right)\right]
+(\xi_1\leftrightarrow \xi_2)\bigg\}\,.
\nonumber
\ea
Here, the following notation was introduced for real-valued distribution functions,
\ba
\widehat T_3(\xi_1,\xi_2)&=&\frac12\left[T_3(\xi_1,\xi_2)-T_3(-\xi_1,-\xi_2)\right]\,,
\nonumber
\\
\widehat
T_7(\xi_1,\xi_2)&=&\frac12\left[T_7(\xi_1,\xi_2)+T_7(-\xi_1,-\xi_2)\right]\,.
\ea
The coefficient functions entering \re{FL-x} and \re{F2-x} have poles at $\xi_1=\xi_2$.
Their regularization is uniquely fixed by going from the moment space in~\re{F-mom1} to the momentum-fraction representation~as
\be
 \frac{1}{(\xi_1-\xi_2)^n}\,\equiv\,
 \lim_{\varepsilon\to 0}\frac{\theta(|\xi_1-\xi_2|-\varepsilon)}{(\xi_1-\xi_2)^n},
\label{reg}
\ee
and it amounts to creating a puncture\footnote{The limits $\varepsilon\to
0$ and $N\to\infty$ do not commute. One has to calculate the moments first and then
take the limit $\varepsilon\to 0$.} on the integration axis at $\xi_1=\xi_2$. We recall that $\xi_q=\xi_1$ and
$\xi_{\bar q}=-\xi_2$ have the meaning of the longitudinal momentum fractions
carried by quark and antiquark, respectively, whereas $\xi_g=\xi_2-\xi_1$ defines
the momentum fraction of a gluon, so that $\xi_q+\xi_{\bar q}+\xi_g=0$.
The singularities of the integrand in~\re{FL-x} and~\re{F2-x} at $\xi_1=\xi_2$
can be attributed to a wee gluon in the initial state with a vanishing momentum
fraction. As a check, it is straightforward to verify that the moments of
\re{FL-x} and \re{F2-x} coincide with~\re{F-mom1}, given the
regularization~\re{reg}.

Obviously, Eqs.~\re{FL-x} and \re{F2-x} have to be in agreement with
\re{F-mom1-asym} as $x\to 1$. To see this, we note first that, owing to the
presence of $\delta(\xi_i-x)$ and $\theta(\xi_i-x)$ terms in the r.h.s.\ of
\re{FL-x} and \re{F2-x}, the momentum fraction of one of the quarks is restricted
to $x\le\xi_i<1$. Performing integration over the momentum fraction of the other
quark, we notice that the dominant contribution comes from the vicinity of the
pole at $\xi_1=\xi_2$. As a consequence, the asymptotic behaviour of the
structure functions \re{FL-x} and \re{F2-x} is driven by the terms most singular
at $|\xi_1-\xi_2|\to 0$. For $F_L(x)$ these terms are
$\sim\theta(\xi_2-x)/(\xi_1-\xi_2)^2$ and for $F_2(x)$ they look like
$\sim\delta(\xi_2-x)/(\xi_1-\xi_2)^2$ and $\sim\theta(\xi_2-x)/(\xi_1-\xi_2)^3$.
Assuming that the non-perturbative distributions $\widehat T_3(\xi_1,\xi_2)$ and
$\widehat T_7(\xi_1,\xi_2)$ have the same behaviour for $\xi_g=\xi_2-\xi_1\to 0$,
we find that the large $x$ asymptotics of \re{FL-x} and \re{F2-x} is governed by
the distribution amplitude $T_7(\xi_1,\xi_2)$.

The asymptotic behaviour of the structure functions can be associated with the properties
of the specific configuration probed at large $x$: it describes the correlations between
a state consisting of a quark carrying  momentum fraction $\xi_1 \sim x$ and a state consisting of a quark and a gluon, carrying  momentum fractions
$\xi_2\sim x$ and $\xi_g=\xi_2-\xi_1\to 0$, respectively.

One can come to the same conclusion in the moment space by examining the properties of
the coefficient functions entering \re{gen-mom}. Assuming that the functions  $T_{\alpha}(\xi_1,\xi_2)$ have a similar, non-singular behaviour, the large $N$ behaviour of $\widetilde{T}_{\alpha}(N)$ is controlled by the large $N$ behaviour of $\Phi_{\alpha,N}(\xi_1,\xi_2)$. One finds that at large $N$ the
functions $\Phi_{3,N}(\xi_1,\xi_2)$ and $\Phi_{7,N}(\xi_1,\xi_2)$, defined in the
region $-1\le \xi_1\,,\xi_2\,,\xi_1-\xi_2 \le 1$, are peaked at the corners of the
phase space $\xi_1=\xi_2=\pm 1$ and, therefore, the moments of the structure
functions $\widetilde{F}_2(N)$ and $\widetilde{F}_L(N)$ receive dominant contribution at large $N$ from
the vicinity of these points.
Comparing the contribution of the two operators in \re{D-moments},
we note that the peak value of $\Phi_{7,N}(\xi_1,\xi_2)$ at  $\xi_1=\xi_2=\pm 1$ is
$\Phi_{7}=N(N^2-1)/6/(N+2)$, which is larger than the corresponding peak of $\Phi_{3,N}=N+1$.
At large $N$, the resulting hierarchy between the two is proportional to $N$, in
agreement with the conclusion we reached examining the ultraviolet divergent contribution in Eq.~\re{UV-hierarchy}.

We conclude that the configuration which dominates the Born-level
twist four contribution at large $x$ includes a gluon with a {\em small} longitudinal momentum fraction $\xi_g=\xi_2-\xi_1\to 0$.
In this way it approaches the twist-two situation where the gluon is absent altogether.
We also saw that the domination of the twist-four operator
${\cal O}_7$ with respect to ${\cal O}_3$ at large~$x$ can be established independently of the ultraviolet dominance assumption.

Finally, let us discuss the reason for the operator
${\cal O}_7$ to be dominant at large $x$ as compared with ${\cal O}_3$.
The difference between the two operators~\re{O37}
manifests itself in the different power of the pre-factor $(py)$ in the r.h.s.\ of \re{T37}.
The factor $(py)$ has the meaning of the light-cone time. For $x$ away from the end-point
region $x=1$ one has $(py)={\cal O}(1)$. On the other hand, for $x\to 1$, or
equivalently $N\to\infty$, one finds that $(py)\sim N$. As a consequence, the
contribution of the operator ${\cal O}_7$ to the moments of the structure
functions is enhanced by a factor $N$ as compared with ${\cal O}_3$.
To show this, it proves convenient to return to \re{F-mom} and
rewrite the moments $\widetilde{D}_i(N+1)$ at large (even) $N$ in the following
form
\be
\widetilde{D}_i(N+1)\equiv \int_0^1d\xi\,\xi^{N}\, D_i(\xi) \approx
\int_0^\infty d\xi\, \e^{-N(1-\xi)}\, D_i(\xi),
\label{Laplace}
\ee
where we have replaced $\xi^N=\exp(N\ln\xi)=\exp(-N\,[1-\xi+\CO((1-\xi)^2)])$, and extended
the integration region by taking into account that $D_i(\xi)=0$ for $\xi>1$.
Using \re{tw4_mat_el}, we express $D_i(\xi)$ as the Fourier transform of the
matrix elements of non-local light-cone operators $Q_i(y)$,
\be
D_i(\xi)=\int_{-\infty}^\infty \frac{d(py)}{\pi}\, \e^{-2i\xi(py)}
\vev{p|Q_i(y)|p}\,.
\label{inv_Fourier}
\ee
Substituting this relation into \re{Laplace} and performing integration over
$\xi$ one obtains,
\be
\widetilde{D}_i(N+1)=\int_{-\infty}^\infty \frac{d(py)}{2\pi i}
\vev{p|Q_i(y)|p}\frac{\exp(-N)}{(py)+iN/2}.
\label{D-pole}
\ee
To evaluate the integral over $(py)$, one uses the fact $\vev{p|Q_i(y)|p}$ has
 no singularities in the lower half plane\footnote{This can be seen from examining Eq.~\re{inv_Fourier} for $\xi>1$, where $D_i(\xi)$ must vanish, so that closing the $(py)$ contour in the lower half plane encloses no poles.}. Then, assuming an appropriate
behaviour of the matrix element $\vev{p|Q_i(y)|p}$ at infinity and closing the
integration contour in~\re{D-pole} in the lower half plane we get the
following remarkable relation~\cite{KM},
\be
\widetilde{D}_i(N+1)=-\vev{p|Q_i(y)|p}\,\e^{-2i(py)}\bigg|_{2(py)=-iN}\,,
\label{D-res}
\ee
which is valid at large $N$.
Here the factor $\exp(-2i(py))$ compensates the
phases coming from the wave functions of quarks separated along on the light-cone
at the large distance $\sim(py)$. Substituting~\re{D-res} into~\re{F-mom}, we
find that the moments of the twist-four structure functions $F_L(x)$ and $F_2(x)$
at large $N$ are determined by the matrix elements of {\it non-local\/} light-cone
operators $\vev{p|Q_i(y)|p}$ analytically continued into the complex plane and evaluated at
at large light-cone separations, $2(py)=-iN$. A similar relation can be established~\cite{KM} for the large $N$ asymptotics of twist two,
\be
q_N=\frac{\vev{p|\overline{\Psi}(y) \ysl \Psi(-y)|p}}
{2(py)}\,\e^{-2i(py)}\bigg|_{2(py)=-iN}\,.
\label{q-tw-2}
\ee
Recall now that the operators
$Q_i(y)$ are given by a linear combination~\re{A12} of twist-four light-cone
operators,~${\cal O}_\alpha$. Comparing the two matrix elements defined
in \re{T37}, we observe that $\vev{p|{\cal O}_7(v;y)|p}$ is enhanced by an
additional power of $(py)$ as compared with $\vev{p|{\cal O}_3(v;y)|p}$.
According to~\re{D-res}, the contribution of the latter operator to the moments
of the twist-four structure functions~$F_L(x)$ and~$F_2(x)$ is suppressed by an
additional factor~$N$ with respect to the former. To arrive at \re{D-moments}
one has to take into account he fact
that the $v$ integrals\footnote{The role of the $v$ integration differs for the case where $|py|$ is large from the expansion in small $(py)$ used to derive
Eq.~\re{Phi37}. It should be noted that the function of $v$ multiplying ${\cal O}_7$ in Eq.~\re{A12} is related to the additional power of~$(py)$ in Eq.~\re{T37} by the requirement that the matrix elements, $\vev{p|Q_i(y)|p}$, should be analytic at small $(py)$.} over the
position of gluon field on the light-cone are dominated by $v \sim 1/(py)\sim
1/N$. This brings the additional factor $1/N$ in front of $\vev{p|{\cal
O}_\alpha|p}$ in agreement with large $N$ behaviour of infrared renormalon contribution~\re{a-37-UV}.

\section{Conclusions and discussion}

\subsection{The simplification of twist four at large $x$ as ultraviolet dominance}

The breakdown of the twist expansion at large~$x$, and, in particular, the
increase in the number of local matrix elements required to parametrize the
contribution of each twist, makes that framework, as it stands, impractical.
Nevertheless, the twist expansion can be used, as we did here at twist four, to
identify the source of the dominant contributions at large~$x$, which scale
as~$\sim N\Lambda^2/Q^2$.

On very general grounds, one expects some simplification in the description of
DIS structure functions at large~$x$. The leading twist simplifies in this limit
since one is essentially probing only the valence parton distribution in the
hadron. On the other hand, in this limit there are large perturbative and {\em
non-perturbative} corrections associated with the formation of a narrow jet in
the final state. Since the formation of the jet proceeds through colour exchange
with the remnants of the target, the same corrections can be associated, by the
OPE, to specific hadronic matrix elements.  The physical picture
we obtained is that the valence parton distribution {\em remains} the relevant
physical quantity at large $x$, in spite of the fact that multi-parton Fock
states give important contributions, which lead to the breakdown of the twist
expansion. The multi-parton correlations measured by individual higher twist
terms reduce in this limit to a trivial part which is associated with the {\em
power-like evolution} of the valence quark distribution.
Since this evolution is a consequence
of the exchange of soft gluons (with momenta~$\sim \sqrt{Q^2(1-x)}$) the finer
structure of the hadron is not resolved.

Our main conjecture in this paper is that the simplification of structure
functions at large~$x$ is realised within the twist expansion through ultraviolet
dominance, namely that each twist is dominated by the contribution which mixes
under renormalization with the leading twist. To prove this statement, it is
necessary to calculate the relevant matrix elements, a task which is definitely
beyond reach of the tools we are using. Nevertheless, examining in some detail
the twist-four contribution at leading order provides significant evidence
supporting this conjecture.

After reviewing the standard twist-two and twist-four analysis we have examined
the fate of infrared renormalons. Following on from~\cite{Beneke,UnPub}, we have
demonstrated that infrared renormalon ambiguities appearing through the
resummation of running-coupling effects in the twist-two coefficient functions
cancel within the OPE, for both $F_2$ and $F_L$, with another ambiguity
associated with the definition of twist-four matrix elements of operators that
mix under renormalization with twist two. While the former ambiguity appears due
to infrared sensitivity of twist two, the latter is directly related to the
ultraviolet renormalization properties of the twist-four operators. In fact,
these are two manifestations of the same phenomenon, namely the arbitrariness in
the separation of different twists. The argument that exact cancellation of the
ambiguity must occur, is equivalent to the statement that the structure functions
must be totally free of factorization scale dependence.

Renormalon ambiguities pose no problem of consistency for the OPE. Full
consistency to order~$1/Q^2$ is achieved upon resumming infrared renormalons at
twist two and including the parameterization of twist four within the same
regularization prescription, e.g. using Borel transforms.

The ultraviolet dominance assumption~\cite{Stein:1996wk}--\cite{KPS} amounts
to a significant simplification: instead of many parameters (a large number of
local matrix elements), the entire twist-four contribution is proportional to the
twist-two matrix elements, and the only non-perturbative parameter is the overall
normalization of this contribution. Under this assumption, the multi-parton
correlation within the target reduces to the trivial part which is proportional
to the twist two matrix elements, whereas any more detailed correlation is
neglected. From the OPE perspective this assumption seems very strong, and it is
probably not justified for general~$x$. On the other hand, the dominant
contributions at large~$x$ are indeed associated with configurations similar to
twist two:
\begin{itemize}
\item{} the recoiling system corresponding to the dominant twist-four contribution to $F_2$~at leading order is that of a {\em single energetic quark}~\re{F-2-x_app}. For~$F_L$ the leading-order recoiling system contains an  energetic quark and an energetic gluon. These are the same configurations that appear at twist two.
\item{} the initial state at twist four contains a gluon in addition to the quark or antiquark (in the amplitude or the complex conjugate amplitude, respectively). However, at large~$x$ this gluon is constrained
 to have a small longitudinal momentum fraction and, therefore, the matrix elements
$\vev{p|{\cal O}_\alpha(v;y)|p}$ are effectively independent of~$v$. Whereas the detailed multi-parton correlation of twist four includes separate dependence on~$v$ and~$y$, its dominant
large-$x$ ingredient depends only on~$py$, just like the twist-two matrix elements.
Consequently the number of {\em relevant} parameters does {\em not} increase at large $x$.
\end{itemize}

Having realized that the dominant contributions at large~$x$ are associated with the twist-two like
configurations, the assumption of ultraviolet dominance becomes natural. We conjecture that the simplification that occurs at large $x$ amounts to the dominance of this contribution which mixes under renormalization with the leading twist.

An additional simplification that occurs at large $x$, is that certain operators, whose matrix element become the largest at large light-cone separations $|py|\sim N$, dominate.
To leading order in the flavour expansion, two operators, ${\cal O}_7$ and ${\cal O}_3$, mix with the leading twist. However, at large $N$ the contribution of ${\cal O}_7$ alone dominates. Beyond the leading order in the flavour expansion, more operators, such as the two-gluon operators
${\cal O}_5$ and ${\cal O}_6$ or the four-quark operators ${\cal O}_1$ and ${\cal O}_2$ may contribute as well.

\subsection{Factorization beyond the leading twist}

Our analysis of infrared renormalons was restricted to leading order in
flavour expansion and, in parallel, we have applied the OPE
\re{T-prod1} to the Born level. It is clearly of interest to extend the analysis beyond this order, particularly since we know from the standard twist-two analysis that multi-gluon emission is important at large $x$.

Higher-order corrections to the twist-two coefficient functions at large $x$ are
well understood~\cite{Sterman}--\cite{KM}, \cite{Sotiropoulos:1999hy}.
They come from three different subprocesses: QCD evolution of the incoming quark
state, $q_N$, hard scattering of the incoming quark off the virtual photon,~$H$,
and the propagation of the narrow quark jet in the final state, $J_N$. Due to
different time scales involved in these three processes, they are
quantum-mechanically incoherent, so their contribution is factorized in the
moment space,
\ba
\label{F=HJq}
\widetilde{F}_2^{\rm tw-2}(N,Q^2)\, &=&\, H_2(Q^2/\mu^2)\, J_2(Q^2/N\mu^2)\, q_N(\mu^2) \nonumber \\
\widetilde{F}_L^{\rm tw-2}(N,Q^2)\, &=&\, \frac1N\, H_L(Q^2/\mu^2)\, J_L(Q^2/N\mu^2)\, q_N(\mu^2),
\ea
with $\mu^2$ being the factorization scale. Here, we have indicated explicitly
the dependence of $H$ and $J$ on the relevant physical scales involved:
the momentum transferred~$Q^2$, and the invariant mass of the final state~$W^2\equiv Q^2(1-x)/x\sim Q^2/N$, respectively. Note that in the longitudinal structure function case we extracted from the hard function the factor $1/N$, so that $H_L(Q^2/\mu^2)$, similarly to $H_2(Q^2/\mu^2)$ is independent of $N$. The $\mu^2$ evolution of $q_N$ is induced by the emission of soft gluons with  momentum $k\sim Q(1-x)\sim Q/N$.
The $\mu$ dependence of the quark distribution $q_N$ at large $N$ follows from the
renormalization properties of the non-local light-cone operator in the r.h.s.
of~\re{q-tw-2}
\be
\frac{d~q_N(\mu^2)}{d\ln\mu^2}= -\Gamma_{\rm cusp}(\alpha_s(\mu^2))\cdot  (\ln N) \cdot \,q_N(\mu^2),
\label{q-EQ}
\ee
with the anomalous dimension $\Gamma_{\rm cusp}(\alpha_s)=\alpha_s C_F/\pi +
{\cal O}(\alpha_s^2)$. This equation coincides with the DGLAP evolution equation
with the matrix of anomalous dimension replaced by their large $N$ asymptotic
behaviour. It is important to notice that at large $N$ the quark distribution
evolves autonomously and does not mix with the twist-two gluon distribution. The
latter will necessary involve the emission of a soft quark into the final state,
so the corresponding contribution is suppressed\footnote{ This  can be also seen from the asymptotics of the DGLAP
evolution kernel $P_{qg}(x) \sim (1-x)^0$ as $x\to 1$.} by a
power of $(1-x)\sim 1/N$. Solving \re{q-EQ} we can
resum large Sudakov (double logarithmic) corrections associated with the infrared
evolution of the quark distribution functions from the hard scale $Q^2$ down to
the scale $Q^2/N^2$, corresponding to the transverse momenta of soft gluons
in the final state.

Another source of Sudakov corrections to the twist-two structure functions in
\re{F=HJq} is the jet function $J$. Requiring that the l.h.s.\ of \re{F=HJq}
should not depend on the factorization scale $\mu$ and taking into account that
the $\mu$ dependence enters into the jet function  only through $\alpha_s(\mu^2)$
and $Q^2/(N\mu^2)$, one can obtain from \re{F=HJq} and \re{q-EQ} the evolution
equation on the jet function~\cite{KS_b}. Solving this equation, one can resum
large Sudakov logarithms to the twist-two structure functions \re{F=HJq}. These
corrections originate from an incomplete cancellation between virtual and real
soft gluon contributions and can be attributed to QCD evolution of the outgoing
quark jet carrying  large energy $\sim Q$ but small invariant mass $\sim
\sqrt{Q^2(1-x)}$. Let us recall~\cite{Sotiropoulos:1999hy} that the jet function depends on the nature of the recoiling hard partons and thus it differs in the case of the longitudinal structure function ($a=L$) with respect to the transverse one ($a=2$).

The factorization of the twist-two structure functions into the product of three
factors, $H,$ $J$ and $q$, corresponding to three different subprocesses is a
general property of QCD dynamics in DIS, which should be valid beyond the
perturbative level. The twist four ingredient which mixes under renormalization
with twist two has its own logarithmically-enhanced higher-order corrections,
which include an evolution factor and a jet-function factor,
\be
\left. \widetilde{F}_2^{\rm tw-4}(N,Q^2)\right\vert_{\UV}
=H_2^{\rm tw-4}(Q^2/\mu^2)\,J_2^{\rm tw-4}(Q^2/N\mu^2)\,\frac{N \Lambda^2}{Q^2}\,q_N^{\rm
tw-2}(\mu^2).
\label{F=HJq_tw4}
\ee
A similar formula (with an overall supression of $1/N$) holds for $\widetilde{F}_L^{\rm tw-4}(N,Q^2)$ .
The contribution \re{F=HJq_tw4} depends, of course, on the prescription used to
regularize the renormalon sum in $J_a$ at twist two~\re{F=HJq}. We do not specify
the regularization prescription here and below. The reader should keep in mind
that this regularization is not associated with the conventional factorization (which is based on dimensional regularization and deals only with the logarithmic divergence), and therefore it is not necessarily associated with the factorization scale $\mu$; it can be introduced for example by the Borel transform as in Eq.~\re{T7UVdom}.
The relation between the
principal value regularization of the Borel transform and the cutoff
regularization of the corresponding momentum integrals has been investigated
in~\cite{Thrust_distribution}. The two differ just by renormalon-type power
corrections of {\em calculable} magnitude.

The next observation is that independence of \re{F=HJq_tw4} of
the factorization scale $\mu^2$ at the logarithmic level implies that
the product of the factors $H$ and $J$ obeys the same evolution equation as the resummed twist-two coefficient function in~\re{F=HJq}. Thus, the coefficient functions in~\re{F=HJq_tw4} differ from those in~\re{F=HJq} only by the initial condition for the evolution
\be
\kappa_1^{(a)}\equiv \frac{ H_a^{\rm tw-4}(Q^2/\mu^2)
\,J_a^{\rm tw-4}(Q^2/N\mu^2)}{H_a(Q^2/\mu^2)\,J_a(Q^2/N\mu^2)},\,\qquad
(a=2,L)
\label{kappa1}
\ee
which is a constant, $\kappa_1^{(a)}=\,{\cal O}(N^0)$, up to corrections
suppressed by powers of $\alpha_s$. We stress that $\kappa_1^{(a)}$ is strictly
independent of the ratios $Q^2/\mu^2$ and $Q^2/(N\mu^2)$. Note that
$\kappa_1^{(a)}$ is not necessarily saturated by the single dressed gluon
contribution, which is strictly associated with the light-cone operator ${\cal
O}_7$. Due to contributions of the two-gluon and the four-quark light-cone
operators, which appear beyond the large $\beta_0$ limit, $\kappa_1^{(a)}$ can
differ from $\kappa_1^{[7]}$ of Eq.~\re{T7UVdom}. In particular, at this level,
the power corrections to $F_2$ and $F_L$ may not be controlled by the same
operators.

In conclusion, we find that ultraviolet dominance implies that the large $N$ asymptotic behaviour of the structure function $F_2$ to twist-four accuracy takes the form,
\ba
\label{F-final}
\widetilde{F}_2(N,Q^2)\,&=&\,\widetilde{F}_2^{\rm tw-2}(N,Q^2)+\widetilde{F}_2^{\rm tw-4}(N,Q^2)+
\cdots
\\\nonumber
\,&=&\,H_2(Q^2/\mu^2)\,J_2(Q^2/N\mu^2)\,q_N^{\rm tw-2}(\mu^2)\left[1+ \kappa_1^{(2)}\,\frac{N\Lambda^2}{Q^2}\right]+\cdots,
\ea
where the ellipses stand for higher-twist contributions, as well as for terms
that are suppressed by $1/N$ at twist two and four. A similar factorization
formula holds for $\widetilde{F}_L(N,Q^2)$. We stress that $\kappa_1^{(a)}$ do
depend on $a=2,L$. From \re{kappa1} and \re{F-final} it follows that
\be
\kappa_1^{(a)}\Lambda^2=\lim_{{N\to\infty}\atop {Q^2/N~{\rm fixed}}}
\frac{\widetilde{F}_a^{\rm tw-4}(N,Q^2)}
{(N/Q^2)\widetilde{F}_a^{\rm tw-2}(N,Q^2)}\,.
\ee
Here the renormalon contribution to the coefficient functions of both twist two
and twist four is regularized. In addition, the ultraviolet divergent
contribution to the twist-four matrix elements is regularized consistently with
the renormalon regularization at twist two.

Going to large $x$ one cannot consider twist four as the only relevant power
correction. As mentioned above, the twist six contribution is expected to scale as
$N^2/Q^4$. This is reflected in the renormalon contribution~\cite{DasW_DIS}.
The analysis of twist six within the light-cone expansion is
significantly more complicated than that of twist four. However, based on the
experienced gained here, we assume that the same picture emerges: amongst the
large number of matrix elements needed to describe twist six contribution,
the dominating contribution will be the one which mixes under renormalization
with the leading twist. In particular, at Born-level for
twist six, the dominant configuration to $F_2$ will be the one containing a single energetic  quark in the final state.
Similarly, $F_L$ at twist six is dominated by a final state containing an energetic quark and an energetic gluon collinear to the outgoing jet, i.e. the same configuration as the lower-twist contributions.

Thus, ``extrapolating'' from the twist-four results, we expect that the dominant
contributions to {\em any} twist is proportional to the twist-two parton distribution $q_N$. This leads to a generalization of the large-$x$ factorization formula~\re{F-final} to any order in
$N\Lambda^2 /Q^2$,
\ba
\label{tw2_mom_mod}
\widetilde{F}_2(N,Q^2)&=& H_2(Q^2/\mu^2)\,J_2(Q^2/N\mu^2) \,J_2^{\rm NP}(N\Lambda^2 /Q^2) \,q_N^{\rm tw-2}(\mu^2)\nonumber \\
\widetilde{F}_L(N,Q^2)&=& \frac1N \, H_L(Q^2/\mu^2)\,J_L(Q^2/N\mu^2) \,J_L^{\rm NP}(N\Lambda^2 /Q^2) \,q_N^{\rm tw-2}(\mu^2),
\ea
with
\be
J_a^{\rm NP}(N\Lambda^2 /Q^2)=1+ \kappa_1^{(a)}\frac{N\Lambda^2}{Q^2} +
\kappa_2^{(a)}\lr{\frac{N\Lambda^2}{Q^2}}^2+ \cdots,
\label{J-exp}
\ee
where $\kappa_n$ are dimensionless non-perturbative parameters representing the
leading contribution of terms with twist $2(n+1)$. $\kappa_1^{(a)}$ was defined
in~\re{kappa1}.  As argued above, the relative suppression of the longitudinal
structure function with respect to the transverse one is not special to the
leading twist, but is rather common to all the terms in the twist expansion (see
e.g.~\re{F-mom1-asym}).

It thus follows from the OPE and the ``ultraviolet dominance'' assumption that
the dominating power corrections in the large $N$ limit can be all resummed into
a shape function of a {\em single} variable $J_a^{\rm NP}(N\Lambda^2 /Q^2)$ that
multiplies $J_a(Q^2/N\mu^2)$. We stress that \re{tw2_mom_mod} is valid up to
perturbative and non-perturbative corrections which are suppressed by $1/N$.

The jet function itself contains now
two factors, a perturbative factor $J_a(Q^2/N\mu^2)$ that can be calculated
simply using the twist-two operators, and a non-perturbative factor $J_a^{\rm
NP}(N\Lambda^2 /Q^2)$ which sums up power corrections to all orders. The
separation between them by regularizing the renormalon sum in $J_a(Q^2/N\mu^2)$
is, of course, arbitrary. However, the product is regularization independent:
both factors together describe the hadronization of the recoiling jet. Since the hadronization process takes place through the exchange of soft gluons with the remnants of the target, the parameters $\kappa_n^{(a)}$, and thus the function $J_a^{\rm NP}(N\Lambda^2 /Q^2)$, do depend\footnote{In ref.~\cite{Sotiropoulos:1999hy} the factor $J_a$ is defined in a target independent way.
On the other hand the power corrections encoded by $J^{NP}_a$
contain information about the matrix elements of higher twist
operators, through the target dependent constants $\kappa_n^{(a)}$.
Thus we cannot interpret  $J^{NP}_a$ as merely the non-perturbative
corrections to $J_a$, but rather as a more general, target dependent
factor which encodes the entire power corrections to the moments of
the structure functions.} on the target.

Our analysis in this paper was restricted to the parity-conserving spin-averaged structure functions $F_2$ and $F_L$. If the ultraviolet dominance conjecture
indeed holds, it is natural to expect that large-$N$ factorization formulae such as \re{tw2_mom_mod} apply to a wider class of structure functions.
In particular, we expect it to apply to the parity-violating structure function $F_3$ as well as the spin structure function $g_1$, which both share~\cite{DasW_DIS} the same leading twist $1/Q^2$ renormalon ambiguity with $F_1\equiv \frac1{2x}(F_2-F_L)$.  Note, however, that in order to identify the relevant twist-four operators and trace down the cancellation of renormalon ambiguities one must first work out the operator expressions for the twist four contribution, similarly to the ones of~\cite{Balitsky:1989bk}, which are not yet available for these structure functions.

Let us try to understand the physical interpretation of the obtained expression
for the structure functions~\re{tw2_mom_mod}. It is convenient to introduce a new
non-perturbative distribution function $j_a(m^2)$ of a {\em single argument}
\cite{Korchemsky:1998ev,KS_thrust},
\be
J_a^{\rm NP}(N\Lambda^2 /Q^2)=\int_0^\infty d m^2 \exp(-Nm^2/Q^2) j_a(m^2)\,.
\label{J-new}
\ee
By the definition, this function is an inclusive distribution describing the
fragmentation of a partonic system with the invariant mass $m^2$ into the hadronic
final state. $j_a(m^2)$ does not depend on the momentum transfer $Q^2$ and, it is, in this sense, a universal distribution function. Nevertheless, it does depend on the target as well as on the renormalon regularization prescription (see below). Inserting
\re{J-new} into \re{tw2_mom_mod} and going over from the moment space to the
momentum-fraction representation\footnote{At the accuracy considered, i.e. when terms suppressed by $1/N$ are neglected $x^{N-1}\simeq e^{-N(1-x)}$, so the Mellin transform is equivalent to a Laplace transform.}, we obtain the following expression for the
structure function with the power corrections included,
\be
F_a(x,Q^2)=\int_x^1d\xi\, F_a^{\rm tw-2}(\xi,Q^2)\,[Q^2\,j_a(Q^2(\xi-x))], \quad \quad\quad \quad (a=2,L),
\label{F-smearing}
\ee
where we assumed that~$j_a(m^2)$ vanishes fast for $m^2\gsim Q^2(1-x)$.
Here $F_a^{\rm tw-2}$ stands for the twist-two structure function in which infrared
renormalon in the resummed coefficient functions have been regularized.

We conclude that the effect of
the power corrections to all orders in $(N/Q^2)$ is to smear the
twist-two structure functions close to the end-point, $x \sim 1$. Whereas
both $F_a^{\rm tw-2}$ and $j_a$ depend on the renormalon regularization prescription, this dependence
cancels out in their convolution in the r.h.s.\ of~Eq.~\re{F-smearing}.

Let us consider the general properties of the distribution function
$j_a(m^2)$.  The OPE based relation~\re{J-exp} leads to a similar expansion of the function for $j_a(m^2)$,
\be
j_a(m^2)=\delta(m^2)+\kappa_1^{(a)}\Lambda^2\delta'(m^2)+\kappa_2^{(a)}\Lambda^4\delta''(m^2)+\cdots.
\ee
This expansion is singular as it goes over the derivatives of the delta function.
Its substitution into \re{F-smearing} leads to the following twist expansion at
large~$x$
\ba
F_a(x,Q^2)&=&F_a^{\rm tw-2}(x,Q^2)-\kappa_1^{(a)}\frac{\Lambda^2}{Q^2(1-x)}
\left[(1-x)\frac{d}{dx}F_a^{\rm tw-2}(x,Q^2)\right]
\nonumber
\\
&+&
\kappa_2^{(a)}\lr{\frac{\Lambda^2}{Q^2(1-x)}}^2
\left[(1-x)^2\frac{d^2}{dx^2}F_a^{\rm tw-2}(x,Q^2)\right]+ \cdots\,,
\ea
where higher-twist corrections are proportional to the derivatives of twist two.
The twist-four correction agrees with the expression obtained in Ref.~\cite{Guo}.%
\footnote{We are grateful to J.~W.~Qiu for drawing to our attention to this paper.}

In order to reconstruct the shape of the distribution $j_a(m^2)$, one has to
resum the whole series in $N/Q^2$ in~\re{J-exp}. Obviously, this will require a
full knowledge of non-perturbative effects, which is far beyond reach.
Nevertheless, based on a general properties of non-perturbative distributions
\cite{Belitsky:2001ij}, we expect that $j_a(m^2)$ should vanish for $m^2<0$,
increase as a power at small $m^2$, approach its maximal value at $m^2\sim
\Lambda^2$ and, then, rapidly decrease at large $m^2\sim Q^2(1-x)$.

Moreover, there is a way to probe the structure of the non-perturbative
distribution $j_a(m^2)$ by Dressed Gluon Exponentiation~\cite{DGE}, deducing an
ansatz for $J_a(m^2)$. Relying on the perturbative treatment of soft and
collinear radiation one can conclude that renormalon related power-corrections
exponentiate as well. This can be systematically established by studying the
large-order behaviour of the Sudakov exponent: it turns out that the coefficients
of sub-leading Sudakov logarithms are enhanced factorially due to infrared
renormalons~\cite{Thrust_distribution}. The enhancement of sub-leading logs
implies that Sudakov resummation with a fixed logarithmic accuracy is
insufficient in the limit under consideration. Instead, one is obliged to resum
{\em all} the logarithms that are factorially enhanced in $J_a$. As usual, when
$J_a$ is calculated perturbatively to power accuracy, a renormalon ambiguity
shows up. This ambiguity is compensated at the non-perturbative level by
$J_a^{\rm NP}$. The fact that the renormalon ambiguity in the Sudakov exponent is
additive is fully consistent with the factorized form of~\re{tw2_mom_mod} we
obtained above.

At the level of a single dressed gluon, the exponent is equal to
the logarithmically-enhanced part in the single dressed gluon result. It therefore has
{\em two} Borel singularities, at $u=1,2$, suggesting the following ansatz for
the non-perturbative jet function~\cite{DGE},
\be
J_a^{\rm NP}(N\Lambda^2/Q^2) =
\exp\left\{-\omega_1^{(a)}\frac{C_F}{\beta_0}\frac{N\Lambda^2}{Q^2}
-\frac12 \omega_2^{(a)}\frac{C_F}{\beta_0}\frac{N^2\Lambda^4}{Q^4}\right\},
\label{C_NP}
\ee
where $\omega_1^{(a)}=-\kappa_1^{(a)}$ and $\omega_2^{(a)}=-\kappa_2^{(a)}+\frac12\left(\kappa_1^{(a)}\right)^2$
determine to the center and the width of the distribution $j_a(m^2)$, respectively.
Clearly, $\omega_{1,2}$ depend on the regularization of the renormalon sum in the resummed coefficient function $J_a$.
It is interesting to note that the same structure of the exponent, and thus the
same ansatz for power corrections in the large-$x$ region, appears also in the
case of the transverse and the total fragmentation functions in $e^+e^-$
annihilation.

Eq.~\re{C_NP} should be regarded as a minimal ansatz for the parametrization of
power corrections at large $x$, since further renormalon ambiguities in $J_a$
will probably appear once the analysis of the Sudakov exponent~\cite{DGE} is
generalized to include more than a single dressed gluon. In this case further
power corrections should be included in the exponent of Eq.~\re{C_NP}. We must
stress, that the suppression of such multi-gluon contributions to the
non-perturbative exponent is not at all obvious, and it remains to be
checked\footnote{Some phenomenological evidence of such suppression was found in
\cite{rho_H}, in the case of the thrust distribution in $e^+e^-$ annihilation,
barring hadron mass corrections.}. On the other hand, a non-perturbative
distribution which is more involved than~\re{C_NP} can only be consistently used
provided that renormalon resummation in $J_a$ at the corresponding level is
performed.

\vskip 30pt
\noindent{\large {\bf Acknowledgements}} \\ \\
One of us (G.K.) would like to thank M.~Beneke and V.~Braun
for illuminating discussions. G.K. and D.R. would like to thank the CERN theory
division for the hospitality while part of this work was done.


\begin{thebibliography}{9}


\bibitem{MRST}
A.~D.~Martin, R.~G.~Roberts, W.~J.~Stirling and R.~S.~Thorne,
``NNLO global parton analysis'', [hep-ph/0201127].

\bibitem{Catani}
S.~Catani {\it et al.}, ``QCD'', [hep-ph/0005025].

\bibitem{Olness}
S.~Kuhlmann {\it et al.},
``Parton densities at high-$x$'', [hep-ph/0007141].

\bibitem{Jaffe:zw}
R.~L.~Jaffe,
``Spin, Twist And Hadron Structure In Deep Inelastic Processes'',
[hep-ph/9602236].

\bibitem{DDKS}
A.~Devoto, D.~W.~Duke, J.~D.~Kimel and G.~A.~Sowell,
{\em Phys. Rev.}  {\bf D30} (1984) 541.

\bibitem{SanchezGuillen:iq}
J.~Sanchez Guillen, J.~Miramontes, M.~Miramontes, G.~Parente and O.~A.~Sampayo,
{\em Nucl. Phys.}  {\bf B353} (1991) 337.

\bibitem{Zijlstra:1992qd}
E.~B.~Zijlstra and W.~L.~van Neerven,
{\em Nucl. Phys.}  {\bf B383} (1992) 525.

\bibitem{JS}
R.~L.~Jaffe and M.~Soldate,
{\em Phys. Rev.}  {\bf D26} (1982) 49.

\bibitem{EFP}
R.~K.~Ellis, W.~Furmanski and R.~Petronzio,
{\em Nucl. Phys.}  {\bf B212} (1983) 29;
{\em Nucl. Phys.}  {\bf B207} (1982) 1.

\bibitem{Jaffe}
R.~L.~Jaffe,
{\em Nucl. Phys.}  {\bf B229} (1983) 205.

\bibitem{Balitsky:1989bk}
I.~I.~Balitsky and V.~M.~Braun,
{\em Nucl. Phys.}  {\bf B311} (1989) 541.

\bibitem{Simula}
S.~Simula,
{\em Phys. Lett.}  {\bf B493} (2000) 325 [hep-ph/0005315].

\bibitem{Schaefer:2001uh}
S.~Schaefer, A.~Schafer and M.~Stratmann,
Phys.\ Lett.\ B {\bf 514} (2001) 284 [hep-ph/0105174].

\bibitem{KPS}
A.~L.~Kataev, G.~Parente and A.~V.~Sidorov, ``Fixation of theoretical ambiguities
in the improved fits to $xF_3$ CCFR data at the next-to-next-to-leading order and
beyond,''
[hep-ph/0106221].

\bibitem{LEKN}
S.~Liuti, R.~Ent, C.~E.~Keppel and I.~Niculescu, ``Perturbative QCD analysis of
local duality in a fixed W**2 framework,'' [hep-ph/0111063].

\bibitem{Sterman} G. Sterman, {\em Nucl. Phys.} {\bf B281} (1987) 310;
D. Appell, P. Mackenzie and G. Sterman, {\em Nucl. Phys.} {\bf B309} (1988)
259.

\bibitem{CT} S. Catani and L. Trentadue, {\em Nucl. Phys.} {\bf B327}
(1989) 323; ibid.~{\bf B353} (1991) 183.

\bibitem{CMW}
S. Catani, G. Marchesini and B.R. Webber, {\em Nucl. Phys.} {\bf B349}
(1991) 635.

\bibitem{KM}
G.~P.~Korchemsky and G.~Marchesini,
{\em Nucl. Phys.}  {\bf B406} (1993) 225 [hep-ph/9210281];
{\em Phys. Lett.}  {\bf B313} (1993) 433.

\bibitem{Beneke}
M.~Beneke,
{\em Phys. Rept.}  {\bf 317} (1999) 1 [hep-ph/9807443];
M.~Beneke and V.~M.~Braun,
[hep-ph/0010208].

\bibitem{UnPub} V.~M.~Braun, unpublished notes.

\bibitem{DMW}
Yu.L. Dokshitzer, G. Marchesini and B.R. Webber,
{\em Nucl. Phys.} {\bf B469} (1996) 93.

\bibitem{Average_thrust}
E.~Gardi and G.~Grunberg,
{\em JHEP} {\bf 9911} (1999) 016, [hep-ph/9908458].

\bibitem{Beneke:1995pq}
M.~Beneke and V.~M.~Braun,
{\em Nucl. Phys.}  {\bf B454} (1995) 253
[hep-ph/9506452].

\bibitem{Stein:1996wk}
E.~Stein, M.~Meyer-Hermann, L.~Mankiewicz and A.~Schafer,
{\em Phys. Lett.}  {\bf B376} (1996) 177
[hep-ph/9601356].

\bibitem{DasW_DIS}
M.~Dasgupta and B.~R.~Webber,
{\em Phys. Lett.}  {\bf B382} (1996) 273 [hep-ph/9604388].

\bibitem{Maul:1997rz}
M.~Maul, E.~Stein, A.~Schafer and L.~Mankiewicz,
{\em Phys. Lett.}  {\bf B401} (1997) 100
[hep-ph/9612300].

\bibitem{Akhoury:1997rt}
R.~Akhoury and V.~I.~Zakharov, ``Renormalon variety in deep inelastic
scattering,'' [hep-ph/9701378].

\bibitem{Guo}
X.~F.~Guo and J.~W.~Qiu, ``The leading power corrections to the structure
functions,'' [hep-ph/9810548].


\bibitem{Mueller:pa}
A.~H.~Mueller,
{\em Phys. Lett.} {\bf B308} (1993) 355.

\bibitem{UV_dom}
V.~M.~Braun,
``Ultraviolet dominance of power corrections in QCD?'',
[hep-ph/9708386].

\bibitem{BBM}
M.~Beneke, V.~M.~Braun and L.~Magnea,
{\em Nucl. Phys.}  {\bf B497} (1997) 297 [hep-ph/9701309].

\bibitem{KS_DY}
G.~P.~Korchemsky and G.~Sterman,
{\em Nucl. Phys.}  {\bf B437} (1995) 415 [hep-ph/9411211]; ``Universality of
infrared renormalons in hadronic cross sections,'' [hep-ph/9505391].

\bibitem{Korchemsky:1998ev}
G.~P.~Korchemsky, ``Shape functions and power corrections to the event shapes,''
[hep-ph/9806537].

\bibitem{KS_thrust}
G.~P.~Korchemsky and G.~Sterman,
{\em Nucl. Phys.}  {\bf B555} (1999) 335, [hep-ph/9902341].

\bibitem{Thrust_distribution}
E.~Gardi and J.~Rathsman,
{\em Nucl. Phys.}  {\bf B609} (2001) 123 [hep-ph/0103217].

\bibitem{rho_H}
E.~Gardi and J.~Rathsman,
``The thrust and heavy-jet mass distributions in the two-jet region'',
[hep-ph/0201019].

\bibitem{DGE}
E.~Gardi,
{\em Nucl. Phys.}  {\bf B622} (2002) 365
[hep-ph/0108222].

\bibitem{Korchemsky:2000kp}
G.~P.~Korchemsky and S.~Tafat,
{\em JHEP} {\bf 0010} (2000) 010 [hep-ph/0007005].

\bibitem{Braun:2001qx}
V.~M.~Braun, G.~P.~Korchemsky and A.~N.~Manashov,
{\em Nucl. Phys.}  {\bf B603} (2001) 69
[hep-ph/0102313];
{\em Nucl. Phys.}  {\bf B597} (2001) 370
[hep-ph/0010128];
{\em Phys. Lett.} {\bf  B476} (2000) 455
[hep-ph/0001130].

\bibitem{Teryaev}
A.~V.~Efremov and O.~V.~Teryaev,
Yad.\ Fiz.\  {\bf 39} (1984) 1517;
\\ O.~V.~Teryaev, ``Twist--3 In Proton Nucleon Single Spin Asymmetries,''
[hep-ph/0102296].

\bibitem{Sotiropoulos:1999hy}
R.~Akhoury, M.~G.~Sotiropoulos and G.~Sterman,
{\em Phys.\ Rev.\ Lett.} {\bf 81} (1998) 3819 [hep-ph/9807330];
``A non-local {OPE} for hard {QCD} processes near the elastic limit,''
[hep-ph/9903442].


\bibitem{KS_b}
G.~P.~Korchemsky and G.~Sterman,
{\em Phys. Lett.}  {\bf B340} (1994) 96 [hep-ph/9407344].

\bibitem{Belitsky:2001ij}
A.~V.~Belitsky, G.~P.~Korchemsky and G.~Sterman,
{\em Phys.\ Lett.}\ B {\bf 515} (2001) 297 [hep-ph/0106308].




\end{thebibliography}
\end{document}